\documentclass[varg,longauth]{aa} 

\usepackage{subfig}
\usepackage{graphicx}
\usepackage{xcolor}
\usepackage{txfonts} 
\usepackage{natbib,twoopt}
\usepackage[breaklinks=False]{hyperref}
\usepackage{booktabs} 
\usepackage{multirow}
\usepackage{natbib}
\hypersetup{
    colorlinks=true,
    linkcolor=blue,
    citecolor=blue,
    urlcolor=blue
}

\bibpunct{(}{)}{;}{a}{}{,}

\makeatletter
\AtBeginDocument{\g@addto@macro\@enddocumenthook{\let\bibcite\NAT@testdef}}
\makeatother

\makeatletter
\newcommandtwoopt{\citeads}[3][][]{\href{http://adsabs.harvard.edu/abs/#3}%
{\def\hyper@linkstart##1##2{}%
\let\hyper@linkend\@empty\citealp[#1][#2]{#3}}}
\newcommandtwoopt{\citepads}[3][][]{\href{http://adsabs.harvard.edu/abs/#3}%
{\def\hyper@linkstart##1##2{}%
\let\hyper@linkend\@empty\citep[#1][#2]{#3}}}
\newcommandtwoopt{\citetads}[3][][]{\href{http://adsabs.harvard.edu/abs/#3}%
{\def\hyper@linkstart##1##2{}%
\let\hyper@linkend\@empty\citet[#1][#2]{#3}}}
\newcommandtwoopt{\citeyearads}[3][][]%
{\href{http://adsabs.harvard.edu/abs/#3}
{\def\hyper@linkstart##1##2{}%
\let\hyper@linkend\@empty\citeyear[#1][#2]{#3}}}
\makeatother

\begin{document}

\title{What’s Missing in AGN Feedback?\\ Lessons learnt from Magneticum, IllustrisTNG and Simba}
\titlerunning{What’s Missing in AGN Feedback?}
\authorrunning{Mazengo et al.}

\author{D.\,T. Mazengo\inst{1,2}\thanks{Daudi.Mazengo@eso.org}
\and P. Popesso\inst{1,3}
\and I. Marini\inst{1,3,7}
\and L. M. Valenzuela\inst{7}
\and N. de Isídio\inst{1}
\and V. Toptun\inst{1}
\and Mirjana Povi\'c\inst{4,5,6}
\and Petri V\"ais\"anen\inst{8,9}
\and J. M. Sunzu\inst{2}
\and R.-S. Remus\inst{7}
\and V. Biffi\inst{10,11}
\and K. Dolag\inst{7,12,3}
\and R. Dav\'e\inst{9,13,14}
\and A. Fraser-McKelvie\inst{1}
\and A. Dev\inst{15}
\and S. Vladutescu-Zopp\inst{7}
\and A. Merloni\inst{16}
\and G. Ponti\inst{16, 17}
\and C. Aydar\inst{16}
\and S. Shreeram\inst{16}
\and L. A. Kahinga \inst{2, 18}
\and J. O. Chibueze\inst{19}
\and P. Privatus\inst{20,21}
}

\institute{
European Southern Observatory, Karl-Schwarzschild-Straße 2, 85748 Garching bei München, Germany
\and The University of Dodoma (UDOM), College of Natural and Mathematical Sciences (CNMS), Department of Physics, 1 Benjamin Mkapa Road, 41218 Iyumbu, Dodoma, Tanzania
\and Excellence Cluster ORIGINS, Boltzmann-Straße 2, 85748 Garching bei München, Germany
\and Instituto de Astrof\'isica de Andaluc\'ia (IAA-CSIC), Glorieta de la Astronom\'ia s/n, 18008 Granada, Spain
\and Astronomy and Astrophysics Department, Entoto Observatory and Research Center (EORC), Space Science and Geospatial Institute (SSGI), P.O. Box 33679, Addis Ababa, Ethiopia
\and Department of Physics, Mbarara University of Science and Technology, P.O. Box 1410, Mbarara, Uganda
\and Universitäts-Sternwarte, Fakultät für Physik, LMU, Scheinerstr. 1, 81679 München, Germany
\and Finnish Centre for Astronomy with ESO (FINCA), FI-20014, University of Turku, Finland
\and South African Astronomical Observatory (SAAO), P.O. Box 9, Observatory 7935, Cape Town, South Africa
\and INAF -- Astronomical Observatory of Trieste, Via Tiepolo 11, I-34143 Trieste, Italy
\and IFPU -- Institute for Fundamental Physics of the Universe, Via Beirut 2, I-34014 Trieste, Italy
\and Max-Planck-Institut für Astrophysik, Karl-Schwarzschild-Str. 1, 85741 Garching bei München, Germany
\and Institute for Astronomy, University of Edinburgh, Royal Observatory, Blackford Hill, Edinburgh EH9 3HJ, UK
\and Department of Physics and Astronomy, University of the Western Cape, Bellville, Cape Town 7535, South Africa
\and International Centre for Radio Astronomy Research, The University of Western Australia, 35 Stirling Highway, Crawley, WA 6009, Australia
\and Max Planck Institute for Extraterrestrial Physics (MPE), Giessenbachstrasse 1, 85748 Garching, Germany
\and INAF -- Osservatorio Astronomico di Brera, Via E. Bianchi 46, 23807 Merate (LC), Italy
\and UNISA Centre for Astrophysics and Space Sciences (U-CASS), College of Science, Engineering and Technology, University of South Africa, Roodepoort, South Africa
\and Department of Astronomy and Astrophysics, University of California Santa Cruz, 1156 High Street, Santa Cruz, CA 95060, USA
\and Department of Natural Sciences, Mbeya University of Science and Technology, Iyunga 53119, Mbeya, Tanzania
\and Department of Physics, Dibrugarh University, Dibrugarh 786004, Assam, India
}

\date{Accepted for publication in Astronomy\&Astrophysics. Received 13 April 2026 / Accepted 13 July 2026}

\abstract
{}
{Accurately balancing gas reservoirs, star formation, and feedback processes across cosmic time remains one of the central challenges for galaxy formation models implemented in modern hydrodynamical simulations. While different feedback prescriptions can reproduce selected local galaxy properties with varying success, some of the most pronounced discrepancies emerge when comparing predictions for the hot gas content of dark matter halos.}
{In this study, we examine three state-of-the-art cosmological simulations: Magneticum, IllustrisTNG, and SIMBA, which struggle to simultaneously reproduce the observed galaxy and halo gas properties in the local Universe. We confront their predictions with spatially resolved galaxy data from MaNGA and recent observational constraints on the hot gas mass fraction--halo mass ($f_{\mathrm{gas}}$--$M_{\mathrm{h}}$) relation derived from eROSITA and Sunyaev--Zel'dovich (SZ) measurements.}
{Reproducing the observed relation $f_{\mathrm{gas}}$--$M_{\mathrm{h}}$ requires strong active galactic nucleus (AGN) feedback. However, such strong feedback often leads to excessive quenching in simulated galaxy populations. Magneticum and SIMBA successfully match the observed gas fraction relation but predict an overabundance of quenched galaxies. In contrast, IllustrisTNG implements comparatively weaker AGN feedback, yielding more realistic star-forming fractions but systematically overpredicting hot gas masses in massive groups and poor clusters.} 
{In general, the stated tensions indicate that the current feedback models remain incomplete, not only in terms of the total energy injected but also in terms of the timing, location and coupling of this energy to the surrounding gas. Our results, therefore, highlight the need to revisit subgrid feedback prescriptions and to develop more self-consistent models capable of simultaneously regulating galaxy growth and the thermodynamic properties of halo gas. Motivated by this discrepancy, a companion study will explore whether the feedback strengths required to match halo gas constraints inevitably lead to overquenching and distorted galaxy demographics.}

\keywords{galaxies: evolution -- galaxies: active -- galaxies: halos -- 
galaxies: clusters: general -- cosmology: theory -- methods: numerical}

\maketitle

\section{Introduction}
Understanding how galaxies and their host halos co-evolve requires a consistent picture of the baryonic cycle across a wide mass range, from Milky Way–sized halos to massive clusters. It is well established that low-mass halos predominantly host star-forming galaxies, while massive halos tend to host quenched systems. This trend suggests that star formation is regulated by a transition from stellar feedback-dominated processes at low masses to AGN-driven quenching at high masses \citep[e.g.,][]{Dekel+86,2004MNRAS.353..713K, 2012MNRAS.427..968H}. In the galaxy group mass range (i.e. $\approx10^{13-14}M_{\odot}$), classical models predicted that galaxy groups retain a large fraction of their baryons in the hot intragroup medium (IGrM). In contrast, recent observational studies including X-ray measurements from eROSITA reveal that galaxy groups with halo masses $10^{12-14}M_{\odot}$ are strongly depleted in hot gas relative to the cosmic baryon fraction \citep{2025A&A...693A.197Z, refId0}. These systems retain only $\approx(20$--$40)\%$ of the expected baryons within $R_{200}$\footnote{$R_{200}$ the radius enclosing a mean density 200 times the critical density of the Universe.}, implying that feedback processes redistribute baryons far beyond the virial region. Independent constraints from the kinetic Sunyaev-Zel'dovich (kSZ) effect also indicate that halo gas profiles are significantly more extended than predicted by standard models \citep{2025PhRvD.112l3507H, 10.1093/mnras/stag1314, 2026ApJ..1003..151S}. In particular, \citet{2026ApJ..1003..151S} combine X-ray gas fractions with kSZ measurements over $13 < \log_{10}(M_{500}/M_{\odot}) < 14$ and infer that strong feedback suppresses the matter power spectrum by $\approx10\pm2\%$ at $k=1\,h\,\mathrm{Mpc^{-1}}$, consistent with baryons being displaced to large radii. Additional support for extended baryon reservoirs comes from fast radio burst (FRB) dispersion measures, which probe the integrated electron column density along extragalactic sightlines \citep{article}.

Despite these observational indications of strong baryon redistribution, cosmological hydrodynamical simulations often struggle to reproduce the low gas fractions inferred for galaxy groups. Most models rely on subgrid prescriptions for black hole accretion and AGN feedback that operate below the resolution scale, introducing uncertainties in the predicted thermodynamic structure of halo gas \citep{2013MNRAS.428.2966P, 2017MNRAS.465.3291W, 2018ApJS..239...22S}. Moreover, simulations are typically calibrated to match the galaxy stellar mass function \citep{Vogelsberger14, Schaye15}, while hot-gas properties are often tuned using X-ray–selected samples biassed toward gas-rich systems \citep{2010MNRAS.406..822M, McCarthy17, Schaye:2023jqv}. Such samples exclude the X-ray-faint populations identified in volume-limited eROSITA surveys \citep{ilaria_lightcone, Popesso2025average}, leading many simulations to overpredict the hot gas content of group-scale halos despite reproducing galaxy statistics near the Milky-Way mass scale \citep{Shreeram_2025}.

Among current simulation suites, the Magneticum simulations provide an alternative approach. Their feedback parameters are chosen to reproduce cluster scaling relations and the Magorrian relation rather than explicitly tuning the stellar mass function \citep{dolag25}. Nevertheless, Magneticum has been shown to reproduce several key observed galaxy properties, including the stellar mass function over a broad stellar-mass range, the mass--metallicity relation, and the cosmic star-formation history \citep{dolag25}. Furthermore, Magneticum reproduces several hot-gas observables, including gas fractions and luminosity--mass relations, in close agreement with eROSITA measurements \citep{Dolag16,dolag25,2025A&A...700A.167T, refId0}. A key goal of this work is to assess whether this success also extends to reproducing the observed distribution of galaxy populations in the SFR--$M_{\star}$ plane, their AGN activity, and their environmental dependence.

This issue is closely tied to the co-evolution of AGN and their host galaxies. The hot gas constitutes the dominant baryonic reservoir in massive halos, and its cooling within the circum-galactic region can provide a sustained fuel supply for star formation. This process is particularly important in galaxy groups and clusters, where gas accretion occurs predominantly in the hot mode. In this regime, central galaxies are no longer fed by cold gas streams from cosmic web filaments, as these are unable to penetrate the hot halo \citep{Dekel_Birnboim}. Consequently, the thermodynamical state of the hot gas and its regulation by AGN feedback play a key role in controlling the star formation activity of massive galaxies.

In this work, we compare three state-of-the-art cosmological hydrodynamical simulations: IllustrisTNG \citep{pillepich,2018MNRAS.475..624N}, SIMBA \citep{2019MNRAS.486.2827D}, and Magneticum \citep{Dolag16,dolag25}. The comparison is motivated by their markedly different levels of agreement with current measurements of the $f_{\mathrm{gas}}$–$M_{\mathrm{h}}$ relation from X-ray and kSZ observations: Magneticum shows the closest agreement, SIMBA provides an intermediate match, while IllustrisTNG exhibits the largest discrepancy \citep{refId0}. These simulations adopt different implementations of AGN feedback, ranging from dual-mode thermal and kinetic feedback in IllustrisTNG \citep{2017MNRAS.465.3291W,pillepich}, jet-driven and X-ray feedback in SIMBA \citep{2019MNRAS.486.2827D}, and a highly efficient thermal feedback model in Magneticum \citep{Dolag16,dolag25}. Taken together, they provide a controlled framework for testing whether models that successfully reproduce the hot gas content of halos can also recover the observed distribution of galaxy properties.

To enable a fair and robust comparison between simulations and observations, we adopt the $f_{\mathrm{gas}}$–$M_{\mathrm{h}}$ relations from \citet{refId0} and \citet{2026ApJ..1003..151S} (see appendix Fig.~\ref{fig:fgas}), and use MaNGA as our primary low-redshift benchmark for galaxy properties. MaNGA provides spatially resolved integral-field spectroscopy for a statistically representative sample of nearby galaxies with stellar masses $M_\star > 10^{10}\,M_\odot$, enabling accurate measurements of star formation rates (SFR) and stellar masses while minimising aperture biases. Its dense sampling of group environments, where the majority of galaxies reside in the local Universe, makes it particularly well suited for comparisons with cosmological simulations.

The structure of this paper is as follows. In Section~\ref{sec:description_obs} we describe the observational data and simulation samples. Section~\ref{sec:MS_offset} examines the distribution of galaxies on the SFR–$M_{\star}$ plane and compares quenching fractions between observations and simulations. Section~\ref{sec:agn_dist} investigates the distribution of AGN activity, while Section~\ref{sec:env_role} explores the environmental dependence of galaxy evolution. Finally, Section~\ref{sec:summary_conclusion} summarises our results and conclusively highlighting their implications for models of baryon cycling in the local Universe.

We assume flat $\Lambda$CDM cosmological parameters: $H_0 = 70\,\mathrm{km\,s^{-1}\,Mpc^{-1}}$, $\Omega_{\rm m} = 0.3$ and $\Omega_\Lambda = 0.7$. To ensure consistency between simulations and observations, all are rescaled to a \citet{2003ApJ...586L.133C} initial mass function (IMF).

\section{Data sample}\label{sec:description_obs}

This section describes both the observational and simulated data sets used in this study. Each dataset is well described in its respective section to classify galaxies in terms of star formation activity, nuclear activity, and environment.

\subsection{The observational dataset: MaNGA sample} \label{sect:obs}

The Mapping Nearby Galaxies at Apache Point Observatory (MaNGA), part of SDSS-IV \citep{2015AJ....150...19L, 2017AJ....154...28B}, is an optical integral field unit (IFU) survey providing spatially resolved spectroscopy for more than 10,000 galaxies in the redshift range $0.01 < z < 0.15$ \citep{2015ApJ...798....7B, 2022ApJS..259...35A}. Its wavelength coverage (3,600--10,300\,\AA) and spatial extent (typically $\geq1.5\,R_e$) enable spatially resolved measurements of stellar mass, star formation rate (SFR), and emission-line diagnostics across galaxy discs \citep{2017AJ....154...86W, Comerford_2020, 2024ApJ...963...53C}.

We use the Pipe3D value-added catalogue for the MaNGA DR17 release \citep{Sanchez2022}. Integrated stellar masses ($M_\star$) are obtained by summing the stellar mass surface density ($\Sigma_\star$) derived from single stellar population (SSP) fits over unmasked spaxels. We also extract SFRs, spectroscopic redshifts ($z$), and luminosity distances ($D_L$). SFR estimates are based on the mean of SSP fits over 10, 32, and 100\,Myr timescales and are consistent with $\mathrm{SFR}_{\mathrm{H\alpha}}$ measurements \citep{Sanchez2022}. AGN identifications are supplemented using the MaNGA-AGN related catalogues of \citet{Comerford_2020,2024ApJ...963...53C}. 

Starting from the initial 10,220 MaNGA sources, we apply a quality cut ($\mathrm{QCFLAG}=0$) following \citet{Sanchez2022}, leaving 9,828 reliable galaxies. To ensure stellar-mass completeness and robust spectroscopic measurements, we further restrict the sample to $\log(M_\star/M_\odot) > 10$ and $z < 0.085$ following \citet{2019MNRAS.490.5285P} and ensuring reliable SDSS redshifts \citep{2015ApJ...801L..29R}. This threshold lies above the regime where MaNGA volume-completeness corrections diverge at $\log(M_\star/M_\odot)\simeq9.75$ \citep{2022ApJ...937..117F}. A recent analysis based on the same Pipe3D dataset confirms the completeness of the sample in this stellar mass range \citep{2026A&A...709A.263D}. Moreover, this mass regime corresponds to the characteristic turnover mass associated with AGN feedback and galaxy quenching \citep{2019MNRAS.490.5285P, 2023MNRAS.519.1526P} suitable for our AGN study.

The final primary sample contains 6,709 galaxies ($\approx68.30\%$ of the 9,828 quality-selected objects). Pipe3D adopts an initial mass function (IMF) by default of Salpeter \citep{2024MNRAS.52710201E}.

\subsubsection{AGN classification and Multi-wavelength catalogue}\label{sec:AGN_class}

AGN feedback studies require identification across the electromagnetic spectrum \citep{Comerford_2020,10.1093/mnras/stag274}, as emission spans the entire electromagnetic spectrum, giving a window of different physics \citep{2017A&ARv..25....2P}. We therefore adopt a multi-wavelength approach to capture diverse accretion modes.
Our catalogue combines MaNGA-based AGN from the public pyPipe3D optical AGN candidate catalogue ($S_{\rm AGN}$; 227; \citealt{2022ApJS..262...36S})\footnote{\url{https://ifs.astroscu.unam.mx/MaNGA/Pipe3D_v3_1_1/tables/AGNs_candidates.csv}}, which identifies AGN candidates using central BPT diagnostics following \citet{2018RMxAA..54..217S}, and a multi-wavelength compilation ($C_{\rm AGN}$) merging \citet{Comerford_2020}\footnote{\url{https://cdsarc.cds.unistra.fr/viz-bin/cat/J/ApJ/901/159}} (406) and \citet[387]{2024ApJ...963...53C}\footnote{\url{https://data.sdss.org/sas/dr17/env/MANGA_AGN/v2_0_1/manga_agn-v2_0_1.fits}}. After cross-matching and removing duplicates, 249 unique AGN remain (77 $S_{\rm AGN}$; 172 $C_{\rm AGN}$; Table~\ref{tab:agn_summary}).

\begin{figure}[h]
\centering
\includegraphics[width=0.5\textwidth]{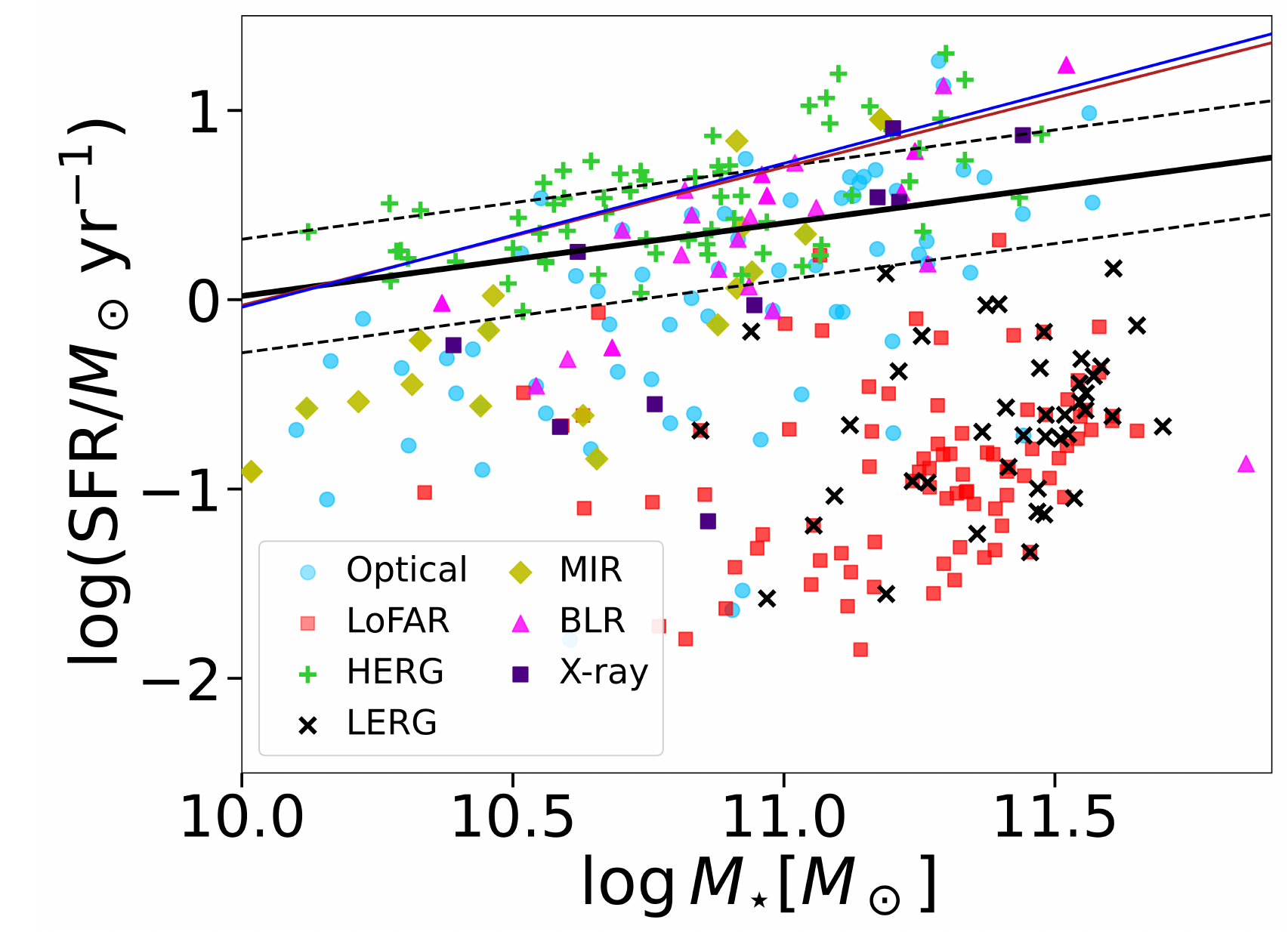}
\caption{Distribution of multi-wavelength AGN in the SFR--$M_\star$ plane (statistics, table~\ref{tab:agn_summary}). Optical AGN (deep sky-blue circles), LoFAR (LoTSS-DR2) AGN as red squares, and NVSS/FIRST radio sources are divided into HERGs (lime green crosses) and LERGs (dark crosses). Additional populations include MIR-selected AGN (lemon green diamonds), broad-line AGN (magenta triangles), and hard X-ray AGN (indigo squares). The solid black line denotes the MS relation from \citet{2019MNRAS.483.3213P} (Po19), with dashed lines marking the $\pm0.3$ dex intrinsic scatter. Alternative calibrations from \citet{2018MNRAS.477.3014B} (BF18) and \citet{2015ApJ...801L..29R} (RP15) are shown in firebrick and blue, respectively.}
\label{fig:plot1}
\end{figure}

To extend sensitivity to low radio luminosities, the MaNGA sample (9,828 galaxies) is cross-matched with the LoTSS DR2 catalogue\footnote{Low Frequency Array (LoFAR) Two-metre Sky Survey \citep{2022A&A...659A...1S}} \citep[][containing more than 4.0 million sources]{hardcastle2023lofar}\footnote{\url{https://cdsarc.cds.unistra.fr/viz-bin/cat/J/A+A/678/A151}} using a $5\arcsec$ matching radius \citep{Mulcahey_2022}. Following \citet{Jin2025}, radio AGN fractions are normalised to the $\sim5{,}500$ MaNGA galaxies within the LoTSS footprint. Within this region, 2,436 galaxies ($\approx44.30\%$) are detected at 144\,MHz. Restricting to 2,252 galaxies with $z<0.1$ minimises contamination from star formation \citep{chilufya2024naturecompactradioloudagn}. Selecting sources $>3\sigma$ above the $L_{144}$--SFR relation (appendix~\ref{app:radioAGN}) yields 261 radio AGN, reduced to 245 after applying our primary selection ($M_\star > 10^{10}\,M_\odot$, $z<0.085$), and to 87 after removing duplicates. The final multi-wavelength catalogue therefore contains 386 unique AGN (Table~\ref{tab:agn_summary}; Fig.~\ref{fig:plot1}; Table~\ref{table:1}).

The classification traces distinct evolutionary stages: High excitation radio galaxies (HERGs) and broad-line AGN lie on or above the star-forming main sequence (MS), while low excitation radio galaxies (LERGs) and LoFAR-selected AGN are predominantly quiescent. At high stellar mass, a vertical gradient in accretion rate emerges along the SFR axis, with high-accretion systems in star-forming hosts and low-accretion systems in quenched galaxies. This agrees with \citet{2015MNRAS.447..110S}, supporting AGN activity regulated primarily by central cold gas supply, with the environment acting indirectly.

\begin{table}[ht]
\caption{Median $\log(M_{\star})$ and $\Delta\log(\mathrm{SFR})$ for the AGN, Fig.~\ref{fig:plot1}.}
\label{table:1}
\centering
\begin{tabular}{l c c}
\hline\hline
Selection & Med $\log(M_{\star}/M_\odot)$ & Med $\Delta\mathrm{log(SFR/M_{\odot}yr^{-1})}$\\
\hline
HERG & 10.84 & +0.18 \\
Broad-Line & 10.96 & +0.07 \\
X-ray & 10.90 & $-$0.01 \\
Optical & 10.89 & $-$0.36 \\
MIR & 10.55 & $-$0.36 \\
LERG & 11.47 & $-$1.16 \\
LoFAR & 11.29 & $-$1.38 \\
\hline
\end{tabular}
\end{table}

\subsubsection{Galaxy environment and halo mass estimates}\label{env}

Environmental classifications for the MaNGA sample are derived from the \citet{yang_galaxy_2007} group catalogue, which utilises a halo-based group finder applied to SDSS galaxies \citep{yang2005}. The algorithm identifies $\sim$470,000 groups, providing central/satellite designations and halo mass ($M_{\rm h}$) estimates based on total stellar mass and $R$-band luminosity. This methodology achieves $\sim$80\% accuracy in group membership and $\sim$95\% in central galaxy identification \citep{ilaria_opticallightcone}.

For the $\sim$30\% of central galaxies without halo masses from the group catalogue; mainly isolated systems or those lacking members brighter than $M_R=-19.5$~mag; we estimate $M_{\rm h}$ using stellar–halo mass relations of \citet{2013ApJ...770...57B,2019MNRAS.488.3143B}. We test representativeness against SDSS in the same redshift range (complete to $M_{\star}\sim10^{10}\,M_{\odot}$). Figure~\ref{fig:dist} shows good agreement: centrals dominate at all masses, increasing with $M_{\star}$, while satellites decline, confirming that MaNGA traces the underlying environmental demographics.

\subsection{The simulated dataset}
We evaluate the Magneticum, IllustrisTNG, and SIMBA suites to test their reproduction of the  observed properties of hot halo gas and galaxy populations within dark matter halos, the multi-wavelength AGN activity trend on the SFR--$M_\star$ plane, and the environmental effects on galaxy evolution. These simulations were selected according to their different ability in hot gas retention across three orders of magnitude of halo mass. According to \citet{refId0}, Magneticum currently shows the strongest alignment with eROSITA stacking results, followed by SIMBA, while IllustrisTNG exhibits larger discrepancies (see also appendix Fig. \ref{fig:fgas}). 

\begin{figure}[h!]
\centering
\includegraphics[width=\hsize]{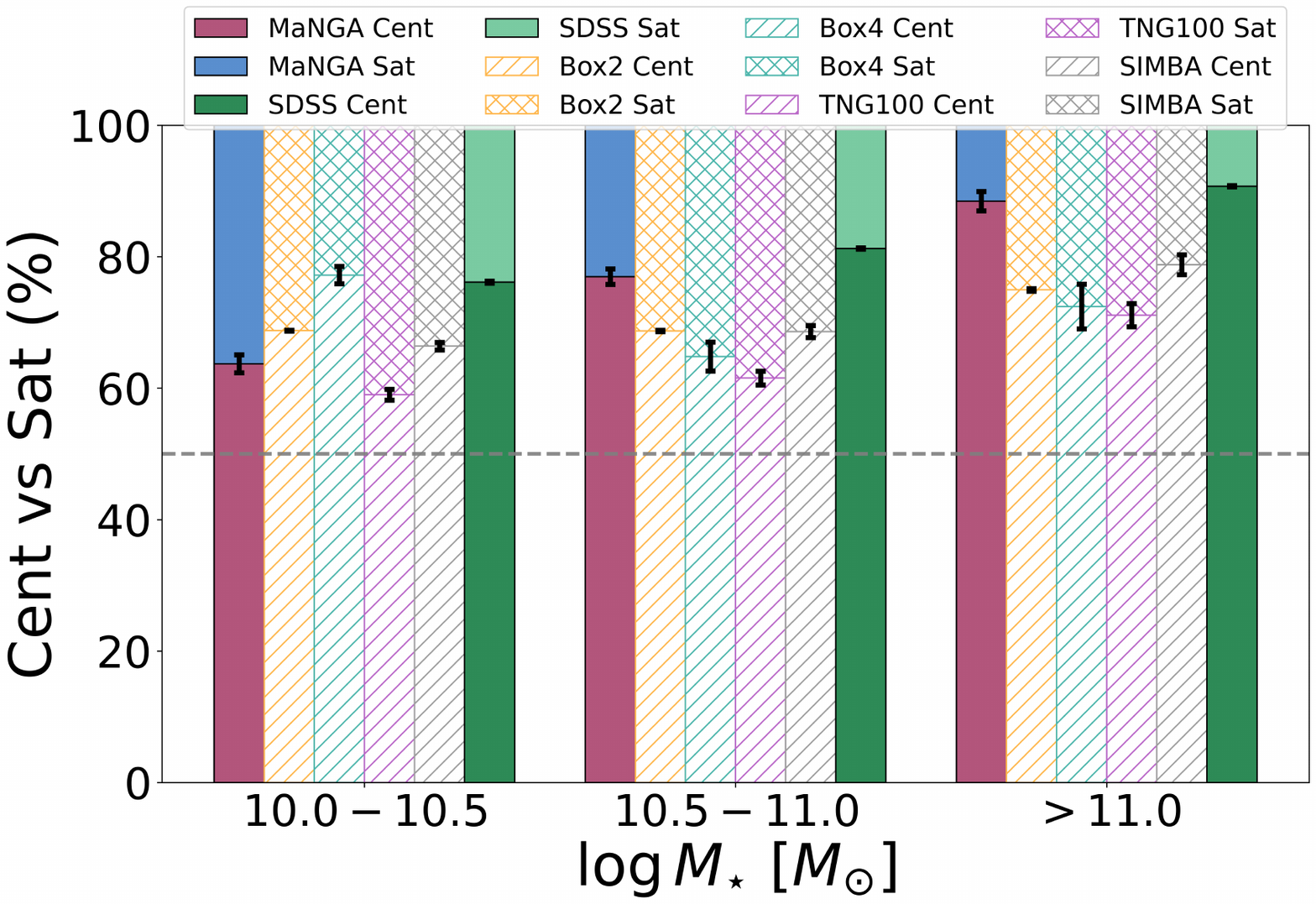}
\caption{Distribution of central and satellite galaxies in our MaNGA-selected sample (purple–blue filled bars), compared to SDSS (green-filled bars) and the simulations considered here. The bars represent the high-resolution (Box4) and medium-resolution (Box2) Magneticum runs, together with IllustrisTNG100 and SIMBA as illustrated in the legend. Black markers with vertical error bars denote the statistical uncertainties on the measured fractions in each stellar mass bin.}
\label{fig:dist}
\end{figure}

To ensure consistency with MaNGA ($z < 0.085$, $M_{\star} > 10^{10}\,M_{\odot}$), we select simulated galaxies with $M_{\star} > 10^{10}\,M_{\odot}$ from $z \approx 0$ snapshots. The SFRs adopted for the simulated galaxies are the catalogue-provided instantaneous SFRs, computed from gas elements that are actively forming stars according to the adopted subgrid star-formation model. Prior to analysing quenching and the scaling relations between $M_{\star}$ and $M_{h}$ across different evolutionary phases, we validate the central--satellite demographics and the consistency between simulations and observations. As shown in Fig.~\ref{fig:dist}, the distributions show good statistical agreement across all datasets.

\subsubsection{Magneticum}
We used two volumes of the Magneticum Pathfinder suite \citet{dolag25} to balance resolution with statistical power. Box4/uhr\citep{2015ApJ...812...29T} (hereafter Box4) ($48\,h^{-1}\,\mathrm{Mpc}$ side) provides the resolution ($m_{\rm gas}=7.3\times10^{6}\,h^{-1}\,M_{\odot}$) to resolve galaxies down to $M_{\star}\approx10^{10}\,M_{\odot}$ \citep{2024A&A...690A.206V}. Box2/hr (hereafter Box2) ($352\,h^{-1}\,\mathrm{cMpc}$ side) samples the massive halos necessary for $f_{\mathrm{gas}}$ constraints. Both adopt a flat WMAP7 $\Lambda$CDM cosmology \citep{2011ApJS..192...18K} and incorporate self-consistent SMBH growth and AGN feedback \citep{hirschmann_cosmological_2014}.

Magneticum successfully reproduces observed galaxy kinematics, angular momentum \citep{2024A&A...690A.206V}, and hot gas fractions \citep{dolag25}. Crucially, both boxes match the central and satellite MaNGA distributions within $1\sigma$ (Fig.~\ref{fig:dist}). For more details of the Magneticum Pathfinder, we refer the reader to\footnote{e.g., \citet{2015MNRAS.451.4277D}, \citet{2015IAUS..309..145R}, \citet{2019trec.confE..26D}, \citet{2022A&A...661A..17B}, \citet{2024A&A...690A.206V}, and \citet{ilaria_erratum}}. This dual-box approach allows us to assess resolution effects while combining the detailed galaxy-scale information from Box4 with the statistical power of Box2. The larger volume of Box2 provides substantially improved sampling of rare massive galaxies ($M_{\star}>10^{11}\,M_{\odot}$). Nevertheless, the inferred central fractions differ by less than $\sim10\%$ between Box2 and Box4 across all stellar-mass bins, and by only $\sim3\%$ in the highest-mass bin (Fig.~\ref{fig:dist}), indicating that the main trends are robust against volume-driven statistical fluctuations.

\subsubsection{IllustrisTNG (TNG100)}
For the IllustrisTNG suite \citep{2018MNRAS.475..624N, 2018MNRAS.475..676S, 2022ApJ...933..161M}, we adopt the TNG100 volume ($\approx110\,\mathrm{Mpc}$ comoving box). This intermediate size provides the optimal combination of volume and resolution ($m_{\rm baryon}=2.1\times10^{6}\,h^{-1}M_{\odot}$) to probe group-scale halos and compare directly with MaNGA demographics \citep{2024MNRAS.532..164L}. The simulation adopts a Planck 2015 $\Lambda$CDM cosmology \citep{2016A&A...594A..16P}.

TNG100 incorporates magnetohydrodynamics and an updated subgrid physics suite, including stellar and AGN-driven feedback \citep{2017MNRAS.465.3291W, 2018MNRAS.473.4077P}. As noted earlier, while TNG's feedback prescriptions yield realistic star-forming fractions, they systematically overpredict hot gas masses in massive groups. This makes TNG100 a critical counterpoint to Magneticum for assessing the trade-off between halo gas retention and galaxy quenching.

\subsubsection{SIMBA}
The SIMBA cosmological simulation \citep{2019MNRAS.486.2827D} is evolved in a $100\,h^{-1}\,\mathrm{cMpc}$ box, providing a balance between mass resolution ($m_{\rm baryon} \simeq 2.68 \times 10^{7}\,h^{-1}\,M_{\odot}$) and cosmological representativeness comparable to TNG100 and Magneticum Box2/hr. SIMBA adopts a Planck 2015 cosmology \citep{2016A&A...594A..16P} and utilises unique subgrid prescriptions for black hole growth, including torque-limited accretion for cold gas and Bondi accretion for hot gas \citep{angles17, 2019MNRAS.486.2827D}.

In SIMBA, the kinetic jet mode is the primary driver for quenching in massive halos, efficiently heating the circumgalactic medium and suppressing accretion \citep{2021MNRAS.500.2036K, 2024MNRAS.534..361S}. As highlighted earlier, SIMBA's aggressive jet-driven feedback successfully matches observed gas fractions but typically results in an overabundance of quenched galaxies, making it an ideal candidate for exploring the physical coupling between halo-scale gas depletion and galaxy-scale overquenching.

\section{Galaxy populations in the SFR-$M_{\star}$ plane}\label{sec:MS_offset}
In this section, we define the SFR-$M_{\star}$ plane loci and use it to classify galaxy populations, which are finally studied by comparing both observation and simulations. 

\subsection{Definition of the loci in the SFR--$M_{\star}$ plane}\label{ms:ms_definition}

To characterise evolutionary pathways and the observed bimodality of the galaxy population \citep[e.g.,][]{2013MNRAS.435.3444P,2025A&A...693A.197Z}, we define the star-forming main sequence (MS) as a reference for classifying galaxies according to their star formation activity. Given our focus on massive systems ($M_{\star} > 10^{10}\,M_{\odot}$), we bypass the low-mass regime of $\log(M_\star/M_\odot)\simeq9.75$ where MaNGA volume-completeness corrections diverge as confirmed by \citet{2022ApJ...937..117F} and adopt the log-linear parameterisation from \citet[Po19]{2019MNRAS.490.5285P} defined by: $\log(\mathrm{SFR}_{\mathrm{MS}})
=
(0.38 \pm 0.04)\,\log M_{\star}\,[M_{\odot}]
-
(3.83 \pm 0.44)$. 

This relation provides a better description of the high-mass MaNGA ridge than steeper relations such as \citet{2015ApJ...801L..29R} and \citet{2018MNRAS.477.3014B}. Although more recent studies \citep[e.g.,][]{2023MNRAS.519.1526P} confirm the bending of the MS at high stellar masses, Po19 remains well suited to the local massive-galaxy regime explored here and captures the high-mass turnover that is central to our analysis.

Our goal is to compare the full distribution of galaxies across the SFR--$M_{\star}$ plane rather than simply separate star-forming and quiescent systems. Since the MS exhibits both a non-zero slope and a mass-dependent turnover, fixed sSFR thresholds can introduce stellar-mass-dependent classification biases, particularly at the massive end where the MS bends. We therefore classify galaxies relative to the MS, which removes the underlying stellar-mass dependence of star formation activity and provides a uniform framework for comparing observed and simulated populations across the full SFR--$M_{\star}$ plane. However, we note that our principal conclusions do not depend on this choice. As a robustness test, we repeated the analysis using a fixed threshold of $\log(\mathrm{sSFR}/\mathrm{yr}^{-1})=-11$ (Appendix~\ref{robust} and Fig.~\ref{fig:sSFR}) and recovered the same qualitative ranking of the simulations. This demonstrates that the trends reported throughout this work are driven by genuine differences in galaxy populations rather than by the adopted classification scheme.

\begin{figure*}
\sidecaption
\centering
\includegraphics[width=0.64\textwidth]{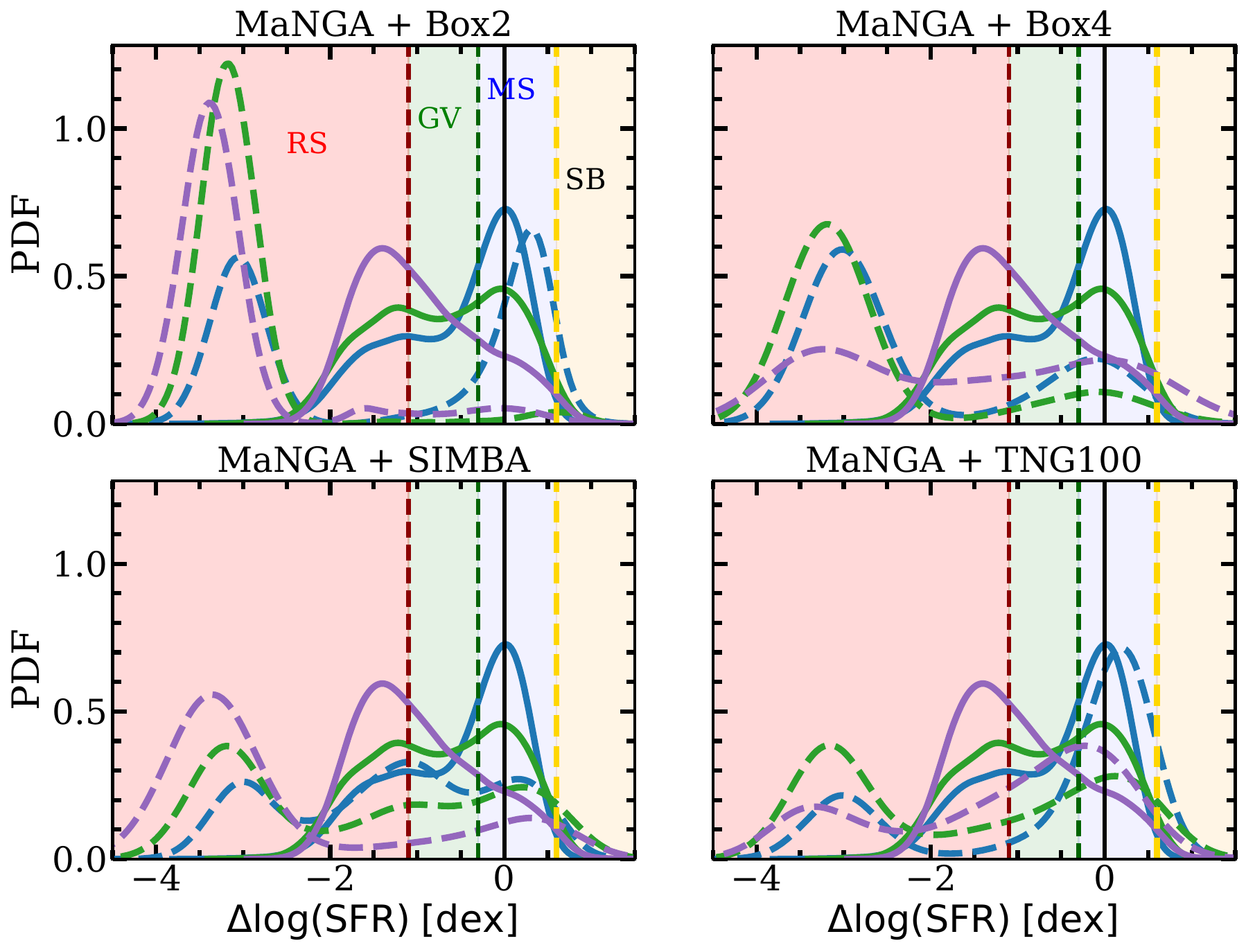}
\caption{
Comparison of MaNGA (solid curves) with simulations (dashed curves) in the SFR--$M_{\star}$ plane, shown as the offset $\Delta\log(\mathrm{SFR})$ [dex] from the Po19 main-sequence relation, for three stellar-mass bins: $10^{10}$--$10^{10.5}\,M_{\odot}$ (blue curve), $10^{10.5}$--$10^{11}\,M_{\odot}$ (green curve), and $>10^{11}\,M_{\odot}$ (purple curve). The top panels compare MaNGA with Magneticum (Box2 left, Box4 right), and the bottom panels compare MaNGA with SIMBA (left) and TNG100 (right). Vertical lines and shaded regions indicate the galaxy populations defined in Sect.~\ref{ms:ms_definition}; the RS, GV, MS, and SB labels shown in the top-left panel apply to all panels. Following \citet{2019MNRAS.486.2827D,2021MNRAS.500.4004D}, simulated galaxies with $\mathrm{SFR}=0$ are assigned random values drawn from a Gaussian centred at $\Delta\log(\mathrm{SFR})\approx-3.0$\,dex for visualisation. The row order reflects decreasing agreement with the observed $f_{\mathrm{gas}}$--$M_{h}$ relation (Appendix, Fig.~\ref{fig:fgas}), linking halo-scale gas depletion to galaxy-scale quenching. Magneticum best reproduces the observed $f_{\mathrm{gas}}$--$M_{h}$ relation, followed by SIMBA, whereas TNG100 provides the closest match to the observed galaxy population in the SFR--$M_{\star}$ plane.
}
\label{fig:MS}
\end{figure*}

By considering the Po19 MS, galaxies in both MaNGA and the simulations are classified according to their offset from the main sequence defined by: $
\Delta\log(\mathrm{SFR})
=
\log(\mathrm{SFR}_{\mathrm{gal}})
-
\log(\mathrm{SFR}_{\mathrm{MS}})$, where $\log(\mathrm{SFR}_{\mathrm{MS}})$ is given by the Po19 MS relation, after rescaling all measurements to a common IMF. Following \citet{2020MNRAS.499..230B}, we define four distinct populations:
\begin{itemize}
    \item Highly star-forming galaxies i.e starbursts (SB): $\Delta\log(\mathrm{SFR}) > 0.6$ dex (pale yellow shaded region from the yellow vertical dashed line in Figure \ref{fig:MS})
    \item Star forming galaxies i.e in the main sequence (MS): $-0.3 < \Delta\log(\mathrm{SFR})< 0.6$ dex (region between yellow and pale green vertical dashed lines in Figure \ref{fig:MS})
    \item Transitional galaxies i.e green valley (GV): $-1.1<\Delta\log(\mathrm{SFR})<-0.3$ dex (pale green shaded region between green and maroon vertical dashed lines in Figure \ref{fig:MS})
    \item Quiescent galaxies i.e galaxies in the red sequence (RS): $\Delta\log(\mathrm{SFR})<-1.1$ dex (the pale red shaded region from the maroon vertical dashed lines in Figure \ref{fig:MS})
\end{itemize}

Figure~\ref{fig:MS} compares the $\Delta\log(\mathrm{SFR})$ probability density functions (PDFs) of MaNGA and the three simulations in three stellar-mass bins: $10^{10}$--$10^{10.5}\,M_{\odot}$ (blue curves), $10^{10.5}$--$10^{11}\,M_{\odot}$ (green curves), and $>10^{11}\,M_{\odot}$ (purple curves). To accommodate finite numerical resolution in simulations and facilitate visualisation, galaxies with unresolved star formation ($\mathrm{SFR}=0$ in simulations) are assigned stochastic values with a Guassian distribution peaked around $\Delta\log(\mathrm{SFR})$= -3 dex from the MS and with a dispersion of 0.3 dex (more information, appendix~\ref{bimodal:ms_bimodality}). While their intrinsic distribution is physically unconstrained, these galaxies would reside significantly below the MS, effectively populating
the quenched region \citep{2019MNRAS.485.4817D}. 

\subsection{Observed versus simulated SFR--$M_{\star}$ plane}\label{sec:galaxy_population}

Direct comparisons between observations and simulations are subject to survey volume, selection effects, and mass-dependent sampling. To mitigate these biases, galaxies are binned in stellar mass (0.5 dex), and we focus on relative distribution shapes rather than absolute normalizations. We analyse the PDF of the offset from the main sequence, where $\Delta\log(\mathrm{SFR}) = 0$ corresponds to the Po19 relation.

As shown in Fig.~\ref{fig:MS}, the MaNGA MS distribution closely follows the Po19 relation, with a typical dispersion of $\approx 0.3$ dex. Toward higher stellar masses, an increasing fraction of galaxies populates the GV and RS. A pronounced accumulation at $\Delta\log(\mathrm{SFR}) \approx -1.5$, just below the GV region, likely reflects a physical or instrumental floor rather than residual star formation (see details, appendix~\ref{bimodal:ms_bimodality}). One of the debates on this apparent “quiescent peak” is that it can be attributed to systematic uncertainties in SFR diagnostics. Infrared tracers may be contaminated by dust heated by old stellar populations \citep{2023A&A...673A..16K}, while UV indicators are affected by emission from evolved stars (e.g., post-AGB and horizontal branch; \citealt{2007ApJS..173..267S, 2016ApJS..227....2S}). In addition, SED fitting and $D4000$ diagnostics suffer from age--metallicity degeneracies and priors that disfavor zero-SFR solutions \citep{2015ApJ...799..125V, 2022ApJ...927..164B}. Together, these effects indicate that the observed quiescent peak primarily reflects methodological limitations rather than ongoing star formation. However, throughout this study, we concur with \cite{2019MNRAS.485.4817D} that these galaxies are attributed to the quiescent region,  and we count them as detailed above. 

\subsubsection{MaNGA versus Magneticum}
Magneticum provides the closest match to the observed hot gas fractions among the simulations considered (Fig.~\ref{fig:fgas}). However, it performs poorly in reproducing the galaxy population, (Fig.~\ref{fig:MS}). The location of the MS is consistent with observations only in the lowest stellar-mass bin ($10^{10.0}$--$10^{10.5}\,M_{\odot}$) when using the lower-resolution Box2 run. No clear green valley (GV) is present: instead, Magneticum produces a single log-normal distribution with a pronounced tail toward SFR=0. This tail hosts a substantial quenched population, highlighted by the artificial Gaussian component placed at $\Delta\log({\rm SFR})=-3$. The resulting overquenching is most severe in the higher stellar-mass bins, where the MS becomes almost indistinguishable.

The comparison with the higher-resolution Box4 run shows a similar overall behaviour. Although Box4 recovers a more visible MS at higher stellar masses-suggesting some resolution dependence-the level of quenching remains excessive (see Table~\ref{table:stats}), with the majority of galaxies still populating the low-SFR tail below the MS. A distinct GV is still absent. Moreover, although MaNGA galaxies at high masses are predominantly quenched, Box4 exhibits an excess of star-forming systems, including in the starburst region (see also Table~\ref{table:stats}).

The persistence of overquenching across different resolutions indicates that this behaviour is driven by the subgrid physics \citep{2024A&A...683A..57R}, rather than by numerical resolution or volume effects. Magneticum systematically overpredicts the quenched fraction, suggesting that its AGN feedback is too efficient at removing or heating the gas, thereby suppressing star formation and preventing galaxy rejuvenation. As a result, the simulated galaxy population is significantly more quenched than observed in the local Universe.

\subsubsection{MaNGA versus SIMBA}
SIMBA is the second-best performer in reproducing the observed hot gas fractions (Fig.~\ref{fig:fgas}). However, the comparison of the $\Delta\log({\rm SFR})$ distributions reveals significant discrepancies with respect to the observed galaxy population (Fig.~\ref{fig:MS}). While SIMBA broadly recovers the location of the MS, the observed MaNGA MS is significantly narrower and more sharply defined, particularly in the $10^{10.0}$--$10^{10.5}\,M_{\odot}$ bin. As in Magneticum (Box4), SIMBA exhibits strong overquenching across all stellar masses. Even in the lowest mass bin, only $\sim23\%$ of galaxies lie on the MS, in stark contrast with the majority of MaNGA galaxies (see Table~\ref{table:stats}). This behaviour is likely driven by the high efficiency of AGN-driven winds and X-ray feedback \citep{2019MNRAS.486.2827D,2024MNRAS.534..361S}.

Unlike Magneticum, SIMBA reproduces the bimodal distribution of galaxies, with a well-defined green valley, particularly at $M_{\star} < 10^{11}\,M_{\odot}$. Nevertheless, most galaxies are still concentrated in the low-SFR tail, as indicated by the prominent peak of the artificial Gaussian distribution at $\Delta\log({\rm SFR})=-3$. We also confirm previous findings of elevated SFRs and a higher incidence of starburst galaxies at low-to-intermediate stellar masses \citep{2011ApJ...739L..40R,2019MNRAS.486.2827D}.

Generally, SIMBA reproduces the main sequence, bimodality, and mass-dependent quenching, but still overquenches galaxies less than Magneticum, yet more than observed.

\subsubsection{MaNGA versus TNG100}
IllustrisTNG exhibits the largest discrepancy with the observed hot gas fractions (Fig.~\ref{fig:fgas}). However, it provides the best agreement with the observed galaxy population among the simulations considered (Fig.~\ref{fig:MS}). Across all stellar mass bins, TNG100 reproduces the MS in good agreement with MaNGA, although it slightly overpredicts the fraction of starburst galaxies ($\sim1.36\%$, Table~\ref{table:stats}), possibly due to the larger hot gas reservoir available for star formation.

As in Magneticum, IllustrisTNG does not show a clearly defined green valley, instead displaying a continuous distribution with a tail toward SFR=0. Despite this, it achieves the closest match to the observed galaxy distribution, particularly in reproducing both the MS and the quiescent population (Table~\ref{table:stats}. A notable discrepancy appears only in the highest stellar mass bin, where an excess of star-forming galaxies suggests that quenching in massive systems is less efficient than observed \citep{2018MNRAS.474.3976G,2023MNRAS.520.5651C}. Tests using the larger-volume IllustrisTNG300 simulation yield consistent results, indicating that these trends are not driven by resolution effects.

IllustrisTNG's weak AGN feedback reproduces the main sequence and a substantial fraction of the quiescent population, consistent with previous studies showing reasonable agreement with observed quenched fractions \citep{2019MNRAS.485.4817D}. However, in our classification scheme, the excess of star-forming galaxies in the highest stellar-mass bin suggests that quenching in the most massive systems is somewhat less efficient than observed, potentially due to the retention of hot halo gas.

Overall, simulations face a key tension: matching CGM thermodynamics leads to overquenching of star formation, while matching galaxy properties fails to remove sufficient gas from the halo.

\section{AGN distribution in the SFR--$M_{\star}$ plane}\label{sec:agn_dist}
AGN feedback has long been proposed as a primary mechanism regulating galaxy evolution, suppressing star formation by heating or ejecting gas from the host \citep{2012A&A...541A.118P,2018RMxAA..54..217S}. In cosmological simulations, this feedback is essential for quenching massive galaxies and shaping black hole-host correlations \cite{2017MNRAS.465.3291W}. However, the location of AGN hosts relative to the star-forming main sequence remains debated \citep{Lammers_2023}, likely due to the episodic nature of accretion and selection biases \citep{2018RMxAA..54..217S}.

\begin{figure*}[htbp!]
  \centering
\includegraphics[width=0.96\textwidth]{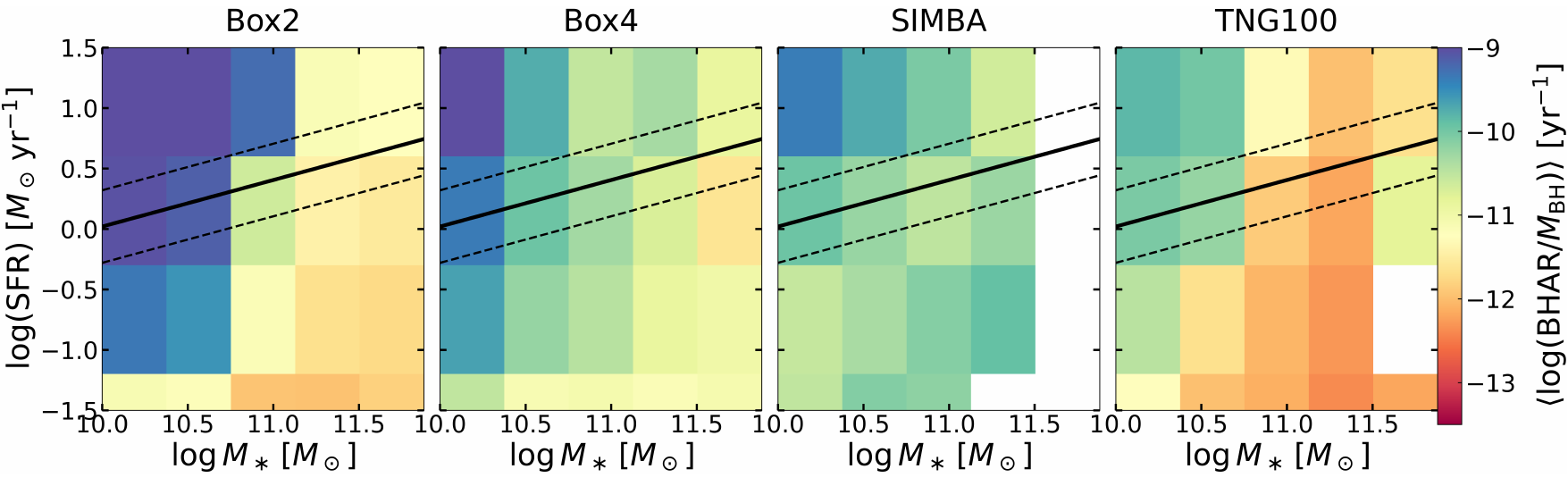}
\caption{SFR-$M_{\star}$ plane in simulations color-coded according to the specific normalized BH accretion, $\log({\rm BHAR}/M_{\rm BH})$ in units of $[{\rm yr}^{-1}]$. From left, the first two panels are for Magneticum (Box2 and Box4, respectively), SIMBA (third panel), and TNG100 (fourth panel). The solid line indicates the location of the Po19 MS, while the dashed line indicates the $1\sigma$ scatter of the relation.}
\label{fig:agn}
\end{figure*}

\subsection{Observation: Accretion gradients traced by SFR}
According to Fig.~\ref{fig:plot1} (see also table~\ref{table:1}), The SFR-stellar mass plane reflects a transition in accretion mode, with declining SFR marking the shift from radiatively efficient to jet-dominated, inefficient feedback, as previously suggested by \citet{Best2012}. This is because most AGN appear regulated by cold gas supply via secular processes \citep{2015MNRAS.447..110S}. 

The radiatively efficient populations - HERGs, broad-line, X-ray, MIR, and optical AGN - predominantly occupy the star-forming regime. HERGs ($\Delta\mathrm{SFR} \approx +0.18\,M_{\odot}\,\mathrm{yr}^{-1}$) trace gas-rich mergers driving high-Eddington accretion and starburst activity \citep{Best2012}. Elevated SFRs in X-ray AGN support this connection, consistent with FIR studies \citep{2017MNRAS.471.3226M, 2019MNRAS.485..452M,2023ApJ...952...12M}.

Broad-line ($\Delta\mathrm{SFR} \approx +0.07\,M_{\odot}\,\mathrm{yr}^{-1}$) and Swift/BAT X-ray AGN ($\Delta\mathrm{SFR} \approx -0.01\,M_{\odot}\,\mathrm{yr}^{-1}$) cluster near the MS, although a few X-ray sources extend below the MS, consistent with \citet{2024A&A...686A..43I} based on eFEDS and LOFAR data. 

Optically selected AGN span a broader evolutionary range (median $\Delta\mathrm{SFR} \approx -0.36\,M_{\odot}\,\mathrm{yr}^{-1}$), tracing the transition from the MS to the RS \citep{2016MNRAS.458L..34E}. While GV occupancy depends on selection \citep{2018RMxAA..54..217S, 2026A&A...706A.376N}, our results indicate that radiatively efficient accretion persists into the GV, consistent with feedback models in which AGN suppress central star formation over Gyr timescales prior to global quenching \citep{2009ApJ...696..891H, 2018RMxAA..54..217S, Lammers_2023}.

Conversely, LERGs ($\log (M_{\star}/M_{\odot}) \approx 11.47$, $\Delta\mathrm{SFR} \approx -1.16\,M_{\odot}\,\mathrm{yr}^{-1}$) and LoFAR-selected AGN ($\log (M_{\star}/M_{\odot}) \approx 11.29$, $\Delta\mathrm{SFR} \approx -1.38\,M_{\odot}\,\mathrm{yr}^{-1}$) trace the endpoint of galaxy evolution. They lie $>1$ dex below the MS and are confined to massive ($\log (M_{\star}/M_{\odot}) > 11.3$), quenched hosts \citep{2025A&A...697A.196I, Jin2025} with old stellar populations \citep{2004MNRAS.353..713K}, indicating a tight connection between radio-mode activity and sustained quenching \citep{Comerford_2020}.

\citet{Jin2025} show that quenching occurred $\approx 5$ Gyr prior to the observed radio phase, implying that jets provide preventative rather than causative feedback. In these systems, kinetic energy dominates the energy budget, offsetting halo cooling and maintaining quiescence \citep[maintenance mode;][]{2025A&A...697A.196I}.

Overall (Fig.~\ref{fig:plot1}), a clear trend emerges at $M_{\star} \approx 10^{11}\,M_{\odot}$: high-accretion AGN lie on or above the main sequence, while low-accretion systems cluster in the quenched regime, producing a strong vertical gradient in accretion rate along the SFR axis. This supports a framework in which AGN activity is primarily governed by central cold gas supply, with environmental effects acting indirectly \citep{2015MNRAS.447..110S}.

\subsection{Simulations: Accretion gradients driven by stellar mass}

We assess whether cosmological simulations reproduce the observed AGN trends in the SFR--$M_{\star}$ plane. Black holes (BHs) are modeled as collisionless sink particles that grow via gas accretion and mergers \citep{2007MNRAS.380..877S, hirschmann_cosmological_2014}, with feedback implemented through subgrid prescriptions that couple accretion energy to the surrounding gas, regulating both BH growth and host evolution \citep{2015ARA&A..53...51S}.

As simulations lack synthetic spectra or broadband luminosities, we use the specific BH accretion rate as a proxy for AGN activity: $\lambda_{\rm BHAR} = \log\!\left(\frac{\mathrm{BHAR}}{M_{\rm BH}}\right)\,[\mathrm{yr}^{-1}]$, where BHAR is the accretion rate and $M_{\rm BH}$ the BH mass. High $\lambda_{\rm BHAR}$ indicates radiatively efficient growth, while low values correspond to maintenance-mode systems. To reduce the influence of poorly resolved low-$M_{\rm BH}$ , we limit the analysis to $10^{6} \leq M_{\rm BH}/M_{\odot} \leq 10^{8.5}$.

\subsubsection{Qualitative agreements: Mass-dependent quenching}

In Fig.~\ref{fig:agn}, all models show a strong dependence of $\lambda_{\rm BHAR}$ on stellar mass ($M_{\star}$), with high accretion rates ($\lambda_{\rm BHAR} > -10.5\,\mathrm{yr}^{-1}$) concentrated in low-mass galaxies, reflecting high gas fractions, though often indicating over-cooling prior to BH seeding \citep{dolag25}. Intermediate accretion ($-11.5 \lesssim \lambda_{\rm BHAR}\,(\mathrm{yr}^{-1}) \lesssim -10.5$) traces systems crossing the green valley.

At $M_{\star} > 10^{11},M_{\odot}$, galaxies are dominated by low accretion ($\lambda_{\rm BHAR} < -11.5,{\rm yr}^{-1}$), consistent with quenched systems. The more gradual decline observed in SIMBA suggests that efficient quenching does not require an abrupt shutdown of black-hole growth. This behaviour is consistent with SIMBA's two-mode accretion and feedback model, in which jet-mode feedback efficiently suppresses star formation while low-level black-hole accretion persists \citep{2019MNRAS.486.2827D}, consistent with the broader distribution of low-to-intermediate accretion states observed in our SIMBA sample. Although SIMBA shows a more gradual decline, all models primarily regulate BH growth with $M_{\star}$ rather than SFR.

\subsubsection{Systematic mismatches: The role of gas supply}

A key discrepancy arises between observations and simulations. MaNGA links black hole accretion to the instantaneous cold-gas supply traced by SFR, whereas simulations regulate AGN primarily via host mass. Consequently, models underpredict high-accretion AGN in massive star-forming galaxies, reflecting radio-mode feedback tied to the most massive systems \citep[see review][]{2015ARA&A..53...51S}. Simulations thus exhibit a dual mismatch: they underpredict high-accretion AGN in massive star-forming galaxies and fail to decouple radio-mode activity from residual star formation in quenched systems.
\begin{figure}[h!]
\centering
\includegraphics[width=\hsize]{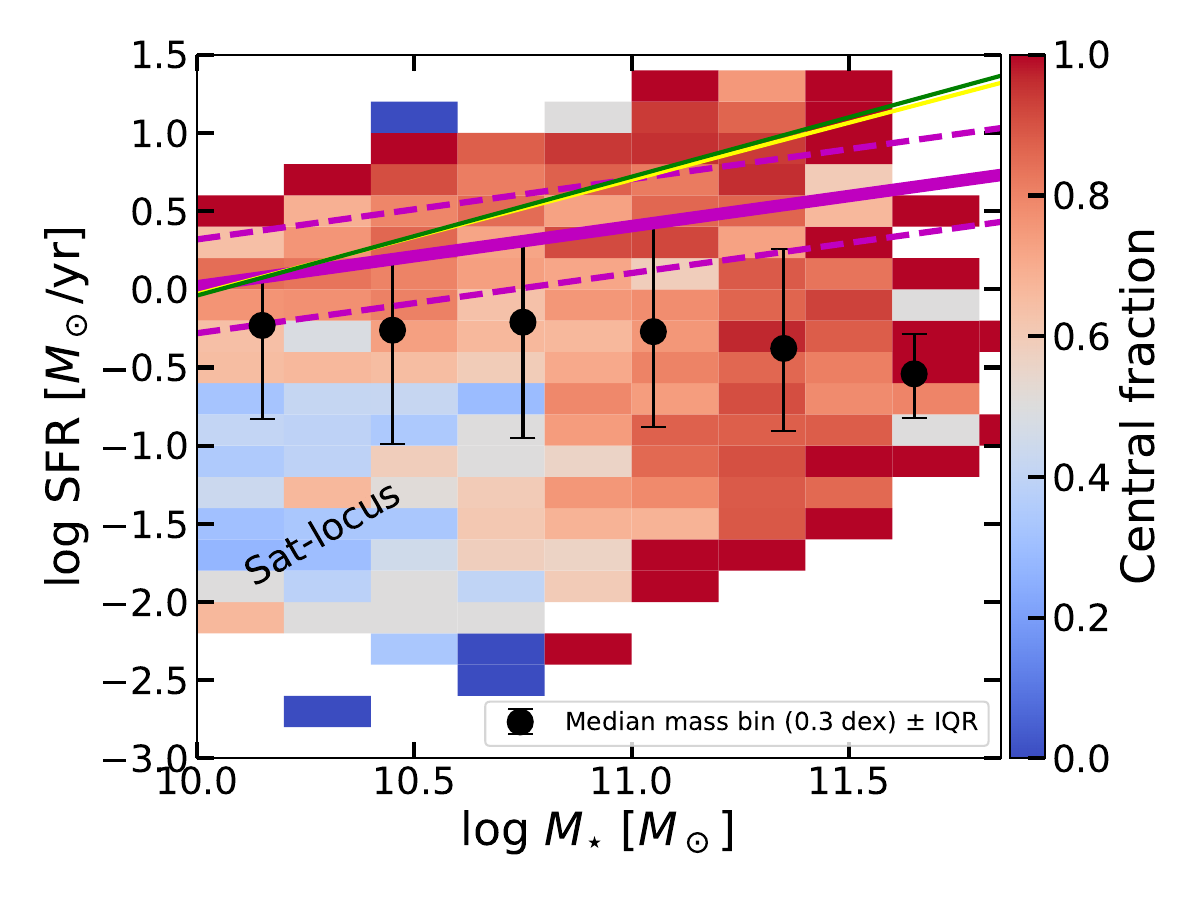}
\caption{Distribution of MaNGA galaxies in the SFR--$M_{\star}$ plane, color-coded by the central fraction (N$_{\rm cent}$/N$_{\rm tot}$) per bin. Solid lines show different star-forming main sequence relations: Po19+$1\sigma$ scatter (solid/dashed magenta), RP15 (yellow), and BF18 (green). Black circles mark the median SFR in 0.3\,dex stellar-mass bins with interquartile range. Centrals ($>60\%$) populate most of the plane except at low masses, while satellites ($<45\%$) concentrate below the MS (satellite locus) at $M_{\star}\lesssim10^{10.8}\,M_{\odot}$.}  
\label{fig:sat}
\end{figure}

\section{Environmental role in quenching}
\label{sec:env_role}

To evaluate environmental dependencies in both datasets, we examine: (i) the centrals and satellites distribution across the SFR–$M_{\star}$ plane, and (ii) galaxy fraction variations across SFR–$M_{\star}$ loci as a function of host halo mass.  In Figure~\ref{fig:sat}, consistent with \citep[e.g.,][as observed in the SDSS sample]{2019MNRAS.483.3213P, 2026A&A...709A.263D}, centrals dominate the MS and high-mass regimes, while satellites are at low-mass ($M_{\star} < 10^{10.5}\,M_{\odot}$) preferentially populating the quiescent region.    

In simulations, central and satellite populations are identified via Friends-of-Friends (FoF) and \textsc{Subfind} algorithms \citep{Springel2001, Dolag2009}, defining the most massive subhalo as the central and the rest are satellites \citep[e.g.,][]{2025MNRAS.538..976M}. Rather than isolating specific physical mechanisms, we only assess whether cumulative environmental prescriptions in Magneticum, IllustrisTNG, and SIMBA reproduce the central-satellite demographics observed in MaNGA.

\begin{figure*}[h!]
\centering
\includegraphics[width=0.8\hsize]{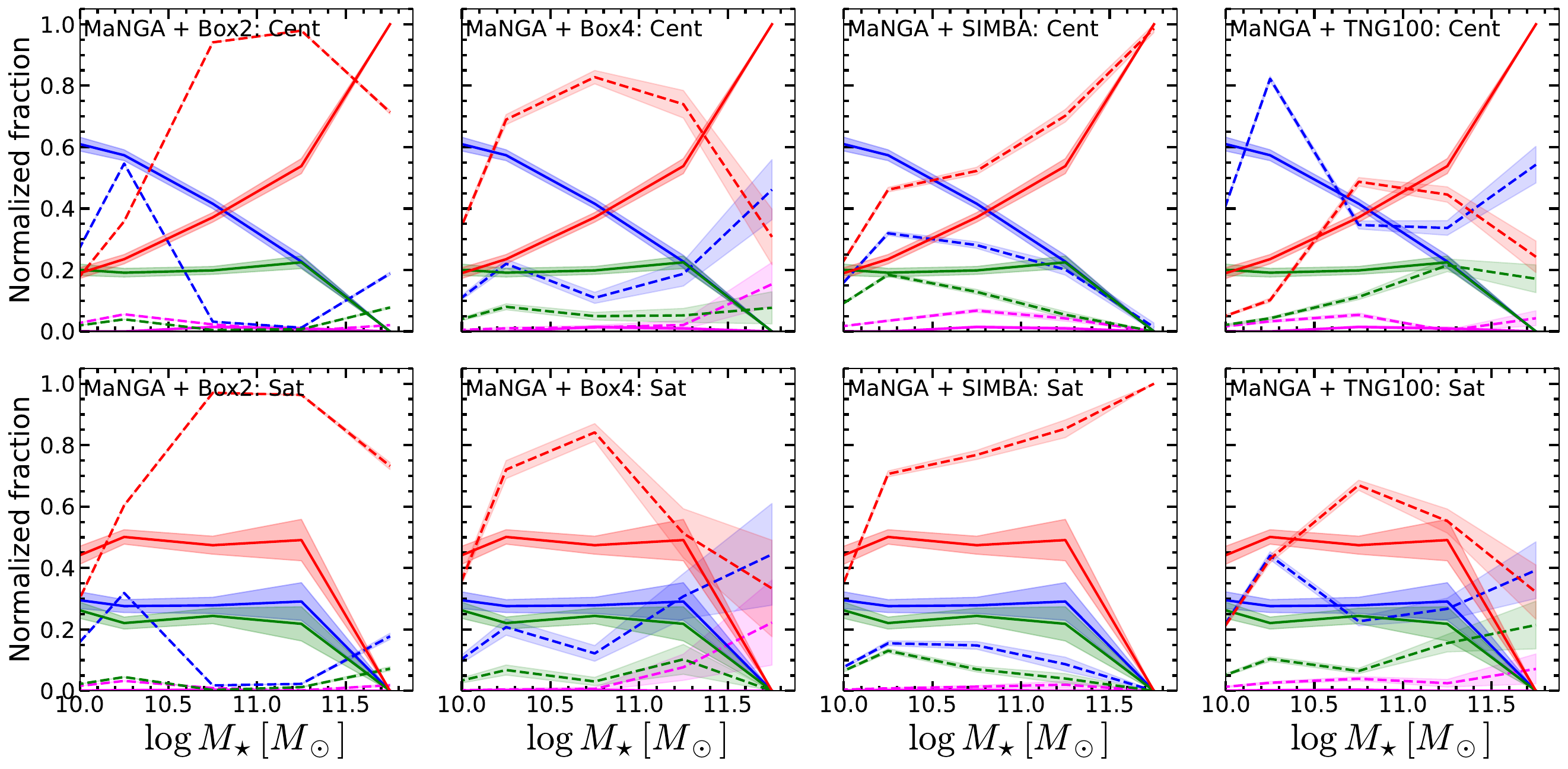}
\caption{Comparison of the galaxy fractions of centrals (top row) and satellites (bottom row) in the different loci of the SFR--$M_{\star}$ plane as a function of stellar mass for MaNGA observations (solid lines) compared with different simulations (dashed lines): Magneticum (Box2-first colum, Box4-second column), SIMBA (third column), and TNG100 (fourth column). In each panel, the curves are color-coded according to the corresponding locus in the SFR--$M_{\star}$ plane, defined relative to the Po19 MS as described in section~\ref{ms:ms_definition}: magenta for starbursts (SB), blue for MS galaxies, green for the green-valley (GV) galaxies, and red for quiescent galaxies (RS). The shaded regions indicate the uncertainties on the fractions. Individual panel titles indicate the sample and simulation combination.}
\label{fig:fractions}
\end{figure*}

\subsection{Stellar mass dependencies for centrals and satellites}\label{sim:env_sim}

Figure~\ref{fig:fractions} shows the fractions of centrals (top row) and satellites (bottom row) as a function of stellar mass across regions of the SFR--$M_{\star}$ plane for MaNGA (solid) and simulations (dashed). In all datasets the MS fraction declines and the RS fraction increases with stellar mass, consistent with mass-quenching trends \citep[e.g.,][]{2010ApJ...721..193P}. Differences in the GV population reflect the non detection of the GV in simulations and the variations in quenching timescales.

\subsubsection{Central galaxies}

MaNGA centrals show a gradual transition from MS-dominated systems at $\log(M_\star/M_\odot)\sim10$ to RS dominance only above $\log(M_\star/M_\odot)\sim11.5$, with a substantial GV population at intermediate masses. 

In Magneticum instead RS fractions dominate by $\log(M_\star/M_\odot)\sim10.5$, while the GV component is strongly suppressed, indicating short transition timescales likely driven by efficient AGN feedback. Retaining residual star formation at intermediate masses may improve agreement with observations \citep[e.g.,][]{2021MNRAS.500.2036K}.

SIMBA shows a similar early transition, with RS fractions exceeding the MS at $\log(M_\star/M_\odot)\approx10.5$ and with a nearly complete quenched population at $\log(M_\star/M_\odot)\approx11.5$. The weak GV population and the negligible SB component are consistent with rapid quenching driven by kinetic AGN feedback and hot-gas suppression of star formation \citep[e.g.,][]{2019MNRAS.486.2827D,2015MNRAS.447..374G}.

IllustrisTNG100 shows the most gradual transition among the simulations. The RS fraction overtakes MS at $\log(M_\star/M_\odot) \sim 10.75$--11.0, still earlier than MaNGA but significantly later than Magneticum or SIMBA. The GV population remains present at intermediate masses, partially overlapping with MaNGA uncertainty bands. For high mass centrals, AGN feedback associated with SMBH growth is likely the dominant internal quenching channel, as suggested by \cite{2021MNRAS.500.4004D}. SB galaxies remain subdominant. This suggests more realistic quenching timescales, in agreement with previous IllustrisTNG analyses \citep[e.g.,][]{2018MNRAS.474.3976G}.

\subsubsection{Satellite galaxies}

MaNGA satellites are more quenched than centrals at fixed stellar mass but still show a gradual transition with a visible GV population.

In Magneticum, satellites are strongly overquenched. RS fractions reach $\sim0.7$--0.9 by $\log(M_\star/M_\odot)\approx10.5$ and approach unity near $\log(M_\star/M_\odot)$ $\approx$ 11, while MS and GV populations nearly vanish. This behaviour is consistent with efficient environmental quenching by hot halo gas \citep[e.g.,][]{2015MNRAS.447..374G}.

In SIMBA, satellites show a similar trend, with RS fractions increasing from $\sim0.7$ at $\log(M_\star/M_\odot)\approx10.25$ and approaching unity toward the highest masses. The GV population remains small and SB galaxies are negligible, indicating strong preventative feedback \citep{2019MNRAS.486.2827D}.

IllustrisTNG100 exhibits the mildest satellite quenching. MS fractions remain significant over $\log(M_\star/M_\odot)\approx10.25$--11.25 and a visible GV component persists at intermediate masses. Although quenched satellites are still somewhat overpredicted at the highest masses, IllustrisTNG100 qualitatively reproduces well the observed pattern of MS, GV, RS and slightly the SB compared to other simulations. Based on our $M_{\star}>10^{10}\,M_{\odot}$ data, the quenched pattern is consistent with results obtained by \cite{2021MNRAS.500.4004D} using the IllustrisTNG100 and SDSS data at $z=0$. 

In summary, simulations broadly reproduce the mass dependence of satellite quenching but differ in strength: Magneticum and SIMBA overpredict quenched satellites, while IllustrisTNG100 provides the closest match to the MaNGA trends.

\subsection{Halo mass dependencies}

Galaxy–halo scaling relations reveal a strong connection between stellar mass and dark matter halo properties \citep[e.g.,][]{Toptun2026}. Although these relations are observationally robust, the physical origin of their scatter remains uncertain and continues to challenge galaxy formation models \citep{dolag25}.

To evaluate the consistency between observations and simulations, we perform two diagnostics. First, we examine the stellar-to-halo mass relation ($M_{cent}/M_{h}$ versus $M_{h}$, hereafter, SHMR). Since environmental processes are primarily governed by the total halo mass of the host potential well in which satellites reside \citep{2013ApJ...770...57B,2010ApJ...710..903M, 2021MNRAS.500.4004D}, we consider only central galaxies. 
The SHMR traces the efficiency of baryon-to-star conversion, which peaks near $M_{h}\sim10^{12}$--$10^{12.5}\,M_{\odot}$ and declines toward higher halo masses \citep{2013ApJ...770...57B,2014ApJ...793...12B,2015MNRAS.450.1604L,2017MNRAS.470..651R,2019MNRAS.488.3143B}. Our analysis focuses on the high-mass regime beyond this peak, where AGN feedback becomes the dominant quenching mechanism \citep{2018AstL...44....8K,2019A&A...631A.175E,2022ApJ...928...28G}. Second, we analyse the satellite fraction across the SFR--$M_{\star}$ plane as a function of halo mass. 
These test provides a direct test of whether feedback and environmental processes prescriptions in hydrodynamical simulations reproduce the stellar content of group halos.

\subsubsection{The stellar-to-halo mass relation}

\begin{figure*}[t]
\centering
\makebox[\textwidth][c]{%
\includegraphics[width=0.9\textwidth]{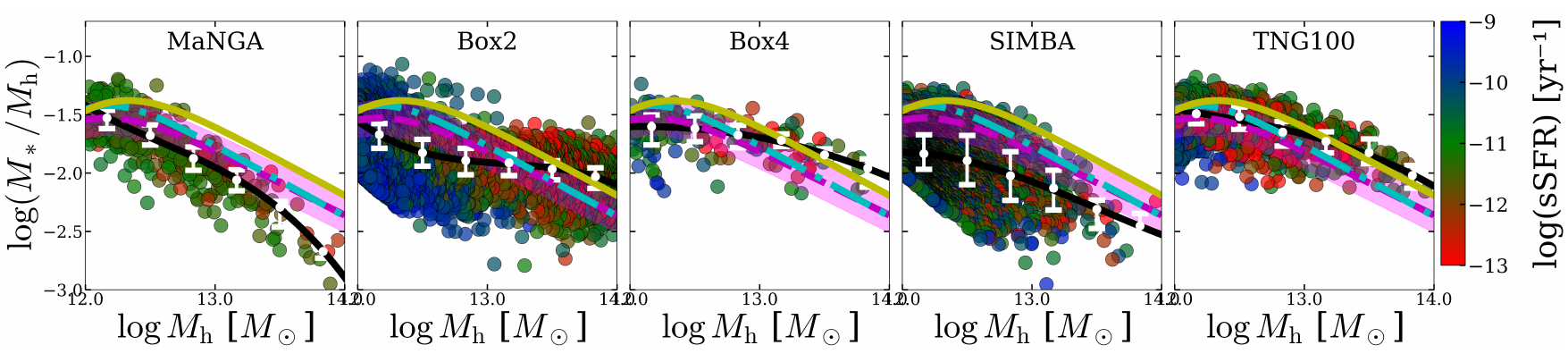}}
    \caption{The stellar-to-halo mass relation ($M_{\star}/M_{h}$ vs.\ $M_{h}$) for MaNGA (first), Magneticum Box2 and Box4 (second and third, respectively), SIMBA (fourth) and IllustrisTNG (fifth). 
Points are color-coded by $\log(\mathrm{sSFR})$. 
Lines show semi-empirical models from \citet{2010ApJ...710..903M} (magenta), 
\citet{2018MNRAS.477.1822M} (yellow), and 
\citet{2013MNRAS.428.3121M} (cyan). 
White points and black curves indicate the median trend.
}
\label{fig:ratio}
\end{figure*}

\begin{figure}[t]
\centering
\includegraphics[width=0.95\columnwidth]{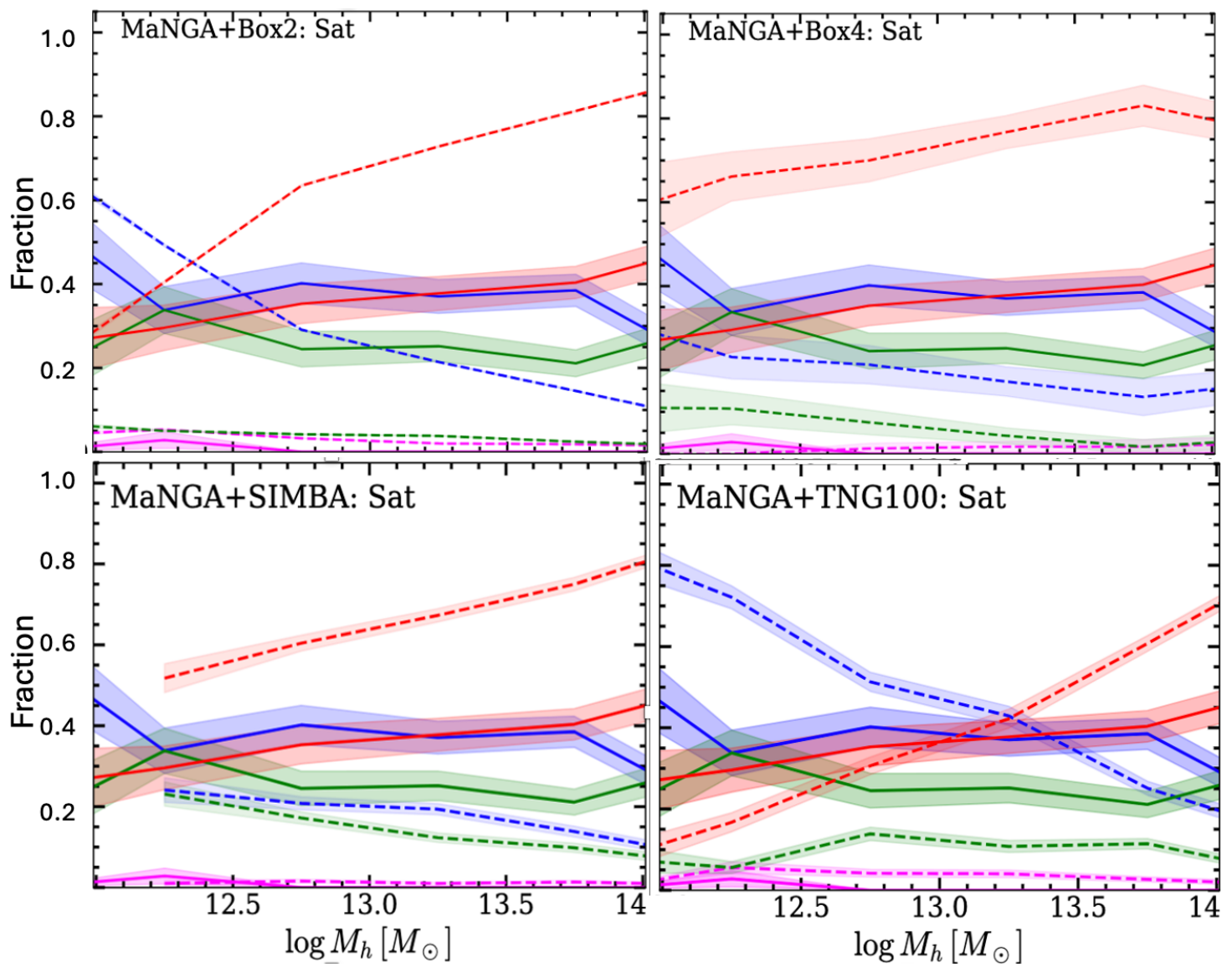}
\caption{Satellite fractions in the SFR--$M_{\star}$ plane as a function of halo mass. Galaxy loci are defined relative to the Po19 main sequence: SB (magenta), MS (blue), GV (green), and RS (red). Solid lines show MaNGA measurements and dashed lines show simulation predictions.}
\label{fig:sat_only}
\end{figure}

Figure~\ref{fig:ratio} shows the SHMR across observations and simulations.
Consistent with previous studies \citep[e.g.,][]{2010ApJ...710..903M,2013MNRAS.428.3121M,2013ApJ...770...57B,2018MNRAS.477.1822M,2019MNRAS.488.3143B,2024ApJ...971...69D,Toptun2026}, the observational data and simulations are broadly consistent with a peak near $M_{\rm h}\approx10^{12}$--$10^{12.5}\,M_{\odot}$, where the SHMR reaches its maximum \citep{2013ApJ...770...57B}. While this feature is less pronounced in Magneticum, particularly Box2, tests including lower-mass haloes indicate that the peak becomes more evident
once the stellar-mass selection is relaxed, suggesting that the weaker feature is primarily driven by the adopted sample
selection. At higher masses, the relation declines, reflecting the transition from cold gas accretion to hot, virialized halo
atmospheres that suppress fresh gas inflow \citep{2019MNRAS.483.3213P}. Our sample is intentionally restricted to
$M_{\star}>10^{10}\,M_{\odot}$, corresponding to the regime where AGN feedback is expected to dominate over stellar feedback and where the tension between reproducing halo gas fractions and galaxy
populations is most pronounced. In this regime, stellar mass growth is increasingly dominated by AGN feedback and ex-situ assembly \citep[e.g.,][]{Toptun2026}. The MaNGA (first panel) SHMR shows mild deviations from semi-empirical models of \citet{2010ApJ...710..903M,2013MNRAS.428.3121M,2018MNRAS.477.1822M}. This likely reflects the limited volume of the MaNGA sample, which underrepresents the most massive brightest cluster galaxies (BCGs). Nevertheless, the median trend remains broadly consistent with \citet{2010ApJ...710..903M}. The scatter does not correlate with log(sSFR) and likely reflects uncertainties in halo mass estimates from the \citet{yang_galaxy_2007} group catalogue.

Magneticum (second and third panels) displays larger differences between the two simulation boxes: Box4 broadly agrees with IllustrisTNG100, whereas Box2 produces a flatter SHMR. The concentration of star-forming galaxies in lower-mass halos in Box2 may result from selection effects or from inefficient early stellar feedback that allows excessive gas cooling \citep{dolag25}. At higher halo masses, continued star formation suggests that AGN feedback may not fully suppress cooling flows.

SIMBA (fourth panel) shows a tighter correlation at low halo masses but produces a lower median SHMR than observed.

IllustrisTNG (fifth panel) shows good agreement with both MaNGA and semi-empirical models, although the simulated relation is slightly flatter. 

Across all simulations, the absence of a strong log(sSFR) dependence in the scatter suggests that local CGM thermodynamic conditions regulate star formation more directly than halo mass alone.

\subsubsection{The role of $M_{h}$ on satellites in the SFR--$M_{\star}$ plane}
Figure~\ref{fig:sat_only} shows the satellite fractions across the SFR–$M_{\star}$ plane versus $M_{h}$. Compared to simulations, MaNGA satellites display a more gradual dependence on $M_{h}$, with fractions nearly constant across the $M_h$ range. 

With the exclusion of the lowest halo mass bin at $M_{h}<10^{12.5}\,M_{\odot}$, Magneticum Box2 and Box4 consistently show a significant overquenching of satellites with 60-80\% of them in the RS. The same is true for SIMBA. 

In contrast, IllustrisTNG100 shows milder quenching, with MS satellites decreasing from (80–40)$\%$ at $M_{h}< 10^{13.}\,M_{\odot}$, to 20\% in the cluster halo regime, where galaxies tend to be more in the RS, consistent with \citet{2021MNRAS.500.4004D}. Overall, while most simulations over-quench satellites, IllustrisTNG100 better reproduces MaNGA trends, whereas SIMBA captures GV systems but quenches satellites earlier than observed.

Despite these differences, all simulations highlight the strong role of the halo mass in satellite quenching, supporting the importance of halo virialization and environmental processes \citep[e.g.,][]{2015MNRAS.447..374G}. The qualitative similar quenched fractions in Box2 and Box4 further suggest sensitivity to subgrid feedback prescriptions.

\section{Summary and conclusions}
\label{sec:summary_conclusion}

This study evaluates whether current feedback frameworks: IllustrisTNG \citep{2018MNRAS.475..624N, 2018MNRAS.475..676S}, Magneticum \citep{dolag25}, and SIMBA \citep{2019MNRAS.486.2827D} can simultaneously reproduce the observed hot-halo gas properties and galaxy demographics. Our comparison with MaNGA reveals a consistent trade-off:

\begin{itemize}
    \item The Magneticum suite (Box2/Box4) effectively matches eROSITA-derived hot gas fractions \citep{dolag25}. However, this successful gas expulsion comes at the cost of severe over-quenching (RS $> 93\%$), causing galaxies to bypass the GV transition phase too rapidly and distorting the observed SFR--$M_{\star}$ plane.
     
    \item SIMBA employs jet-dominated feedback \citep{2019MNRAS.486.2827D} to achieve halo gas depletion. While it moderately reproduces gaseous observations, quenching is triggered prematurely at $M_{\star} < 10^{10.5}\,M_{\odot}$, leading to a faster-than-observed migration toward the RS.
     
    \item IllustrisTNG100, calibrated primarily on galaxy statistics such as the stellar mass function and morphology \citep{2018MNRAS.475..624N, 2018MNRAS.475..676S}, faithfully reproduces MS slopes and observed quenched fractions (e.g., matching the $28\%$ GV fraction). Nevertheless, it lacks sufficient energetic feedback to match eROSITA measurements, retaining too much hot gas in group and cluster-scale halos and maintaining an excess of MS galaxies in all stellar mass bins.
\end{itemize}

We also identify a fundamental mismatch in AGN triggering: MaNGA shows a vertical accretion-rate gradient (SFR-driven), while simulations produce a horizontal gradient (stellar mass-driven), tying black hole growth to host stellar mass rather than star-forming state and availability of gas to fuel the accreation process.

Overall, no current simulation simultaneously reproduces both detailed galaxy demographics and hot-halo gas properties, underscoring the persistent tension between matching stellar and gaseous observables. Achieving a self-consistent model therefore requires feedback that regulates star formation while simultaneously controlling the thermodynamic state of the circum- and intra-galactic medium, something missing in current model prescriptions.

As a way forward, in a companion paper (Paper II) we will test
these conclusions using the FLAMINGO fiducial and extreme
$f_{\rm gas}-8\sigma$ models, investigating whether the feedback strength
required to match halo gas constraints necessarily results in
overquenched or otherwise distorted galaxy populations.

\begin{acknowledgements}

The authors thank the referee for the careful and constructive review of the manuscript. Insightful comments and suggestions helped improve the clarity, presentation, robustness, and overall quality of this work.

DTM acknowledges the administrative and financial support for his PhD studentship provided by the European Southern Observatory (ESO). DTM also deeply acknowledges the tremendous contributions through discussions and/or comments from the following: G. Jin, J. Comparat, A. Mahoro, Y. S. Idala, G. S. Ilha, J. F. Muntama, G. F. Conçalves, Yi Zhang, J. Manyahi, P. V. K. Rao, N. T. Jiwaji, J.D Mbarubucyeye, P. N. Mwaniki, B. A Moltote, S. Sichone, and P. Nkundabakura. PP, VT, and IM have received funding from the European Research Council (ERC) under the European Union’s Horizon Europe research and innovation programme ERC CoG (Grant agreement No. 101045437). KD acknowledges support by the COMPLEX project from the ERC under the European Union’s Horizon 2020 research and innovation program (Grant agreement No. ERC-2019-AdG 882679). MP acknowledges funding from the Spanish MCIU (projects PID2022-140871NB-C21 and PID2024-162972NB-I00), the Severo Ochoa grant CEX2021-515001131-S (MCIN/AEI), and support from the Space Science and Geospatial Institute (SSGI), Ethiopian Ministry of Innovation and Technology.

The calculations for the Magneticum simulations were carried out at the Leibniz Supercomputer Center (LRZ) under project pr83li. This project makes use of the MaNGA-Pipe3D data products; we thank the IA-UNAM MaNGA team for creating this catalogue and the Conacyt Project CB-285080 for their financial support.
\end{acknowledgements}

\bibliographystyle{aa}
\bibliography{paper}

@article{yang_galaxy_2007,
	title = {Galaxy {Groups} in the {SDSS} {DR4}. {I}. {The} {Catalog} and {Basic} {Properties}},
	volume = {671},
	issn = {0004-637X},
	url = {https://ui.adsabs.harvard.edu/abs/2007ApJ...671..153Y},
	doi = {10.1086/522027},
	urldate = {2023-08-24},
	journal = {The Astrophysical Journal},
	author = {Yang, Xiaohu and Mo, H. J. and van den Bosch, Frank C. and Pasquali, Anna and Li, Cheng and Barden, Marco},
	month = dec,
	year = {2007},
	note = {ADS Bibcode: 2007ApJ...671..153Y},
	pages = {153--170},
}

@ARTICLE{ilaria_lightcone,
       author = {{Marini}, I. and {Popesso}, P. and {Lamer}, G. and {Dolag}, K. and {Biffi}, V. and {Vladutescu-Zopp}, S. and {Dev}, A. and {Toptun}, V. and {Bulbul}, E. and {Comparat}, J. and {Malavasi}, N. and {Merloni}, A. and {Mroczkowski}, T. and {Ponti}, G. and {Seppi}, R. and {Shreeram}, S. and {Zhang}, Y.},
        title = "{Detecting galaxy groups populating the local Universe in the eROSITA era}",
      journal = {\aap},
         year = 2024,
        month = sep,
       volume = {689},
          eid = {A7},
        pages = {A7},
          doi = {10.1051/0004-6361/202450442},
archivePrefix = {arXiv},
       eprint = {2404.12719},
 primaryClass = {astro-ph.CO},
       adsurl = {https://ui.adsabs.harvard.edu/abs/2024A&A...689A...7M}
}

@ARTICLE{ilaria_erratum,
       author = {{Marini}, I. and {Popesso}, P. and {Lamer}, G. and {Dolag}, K. and {Biffi}, V. and {Vladutescu-Zopp}, S. and {Dev}, A. and {Toptun}, V. and {Bulbul}, E. and {Comparat}, J. and {Malavasi}, N. and {Merloni}, A. and {Mroczkowski}, T. and {Ponti}, G. and {Seppi}, R. and {Shreeram}, S. and {Zhang}, Y.},
        title = "{Detecting galaxy groups populating the local Universe in the eROSITA era (Corrigendum)}",
      journal = {\aap},
         year = 2025,
        month = mar,
       volume = {695},
          eid = {C1},
        pages = {C1},
          doi = {10.1051/0004-6361/202553853e},
       adsurl = {https://ui.adsabs.harvard.edu/abs/2025A&A...695C...1M}
}

@ARTICLE{pillepich,
       author = {{Pillepich}, Annalisa and {Nelson}, Dylan and {Hernquist}, Lars and {Springel}, Volker and {Pakmor}, R{\"u}diger and {Torrey}, Paul and {Weinberger}, Rainer and {Genel}, Shy and {Naiman}, Jill P. and {Marinacci}, Federico and {Vogelsberger}, Mark},
        title = "{First results from the IllustrisTNG simulations: the stellar mass content of groups and clusters of galaxies}",
      journal = {\mnras},
         year = 2018,
        month = mar,
       volume = {475},
       number = {1},
        pages = {648-675},
          doi = {10.1093/mnras/stx3112},
archivePrefix = {arXiv},
       eprint = {1707.03406},
 primaryClass = {astro-ph.GA},
       adsurl = {https://ui.adsabs.harvard.edu/abs/2018MNRAS.475..648P}
}

@ARTICLE{Daves_24,
       author = {{Popesso}, P. and {Biviano}, A. and {Bulbul}, E. and {Merloni}, A. and {Comparat}, J. and {Clerc}, N. and {Igo}, Z. and {Liu}, A. and {Driver}, S. and {Salvato}, M. and {Brusa}, M. and {Bahar}, Y.~E. and {Malavasi}, N. and {Ghirardini}, V. and {Robotham}, A. and {Liske}, J. and {Grandis}, S.},
        title = "{The X-ray invisible Universe. A look into the haloes undetected by eROSITA}",
      journal = {\mnras},
         year = 2024,
        month = jan,
       volume = {527},
       number = {1},
        pages = {895-910},
          doi = {10.1093/mnras/stad3253},
archivePrefix = {arXiv},
       eprint = {2302.08405},
 primaryClass = {astro-ph.CO},
       adsurl = {https://ui.adsabs.harvard.edu/abs/2024MNRAS.527..895P}
}

@ARTICLE{yang2005,
       author = {{Yang}, Xiaohu and {Mo}, H.~J. and {van den Bosch}, Frank C. and {Jing}, Y.~P.},
        title = "{A halo-based galaxy group finder: calibration and application to the 2dFGRS}",
      journal = {\mnras},
         year = 2005,
        month = feb,
       volume = {356},
       number = {4},
        pages = {1293-1307},
          doi = {10.1111/j.1365-2966.2005.08560.x},
archivePrefix = {arXiv},
       eprint = {astro-ph/0405234},
 primaryClass = {astro-ph},
       adsurl = {https://ui.adsabs.harvard.edu/abs/2005MNRAS.356.1293Y}
}

@ARTICLE{hirschmann_cosmological_2014,
       author = {{Hirschmann}, Michaela and {Dolag}, Klaus and {Saro}, Alexandro and {Bachmann}, Lisa and {Borgani}, Stefano and {Burkert}, Andreas},
        title = "{Cosmological simulations of black hole growth: AGN luminosities and downsizing}",
      journal = {\mnras},
         year = 2014,
        month = aug,
       volume = {442},
       number = {3},
        pages = {2304-2324},
          doi = {10.1093/mnras/stu1023},
archivePrefix = {arXiv},
       eprint = {1308.0333},
 primaryClass = {astro-ph.CO},
       adsurl = {https://ui.adsabs.harvard.edu/abs/2014MNRAS.442.2304H}
}

@ARTICLE{2007MNRAS.380..877S,
       author = {{Sijacki}, Debora and {Springel}, Volker and {Di Matteo}, Tiziana and {Hernquist}, Lars},
        title = "{A unified model for AGN feedback in cosmological simulations of structure formation}",
      journal = {\mnras},
         year = 2007,
        month = sep,
       volume = {380},
       number = {3},
        pages = {877-900},
          doi = {10.1111/j.1365-2966.2007.12153.x},
archivePrefix = {arXiv},
       eprint = {0705.2238},
 primaryClass = {astro-ph},
       adsurl = {https://ui.adsabs.harvard.edu/abs/2007MNRAS.380..877S}
}

@ARTICLE{2010MNRAS.406..822M,
       author = {{McCarthy}, I.~G. and {Schaye}, J. and {Ponman}, T.~J. and {Bower}, R.~G. and {Booth}, C.~M. and {Dalla Vecchia}, C. and {Crain}, R.~A. and {Springel}, V. and {Theuns}, T. and {Wiersma}, R.~P.~C.},
        title = "{The case for AGN feedback in galaxy groups}",
      journal = {\mnras},
         year = 2010,
        month = aug,
       volume = {406},
       number = {2},
        pages = {822-839},
          doi = {10.1111/j.1365-2966.2010.16750.x},
archivePrefix = {arXiv},
       eprint = {0911.2641},
 primaryClass = {astro-ph.CO},
       adsurl = {https://ui.adsabs.harvard.edu/abs/2010MNRAS.406..822M}
}

@ARTICLE{ilaria_opticallightcone,
       author = {{Marini}, I. and {Popesso}, P. and {Dolag}, K. and {Bravo}, M. and {Robotham}, A. and {Tempel}, E. and {Li}, Q. and {Yang}, X. and {Csizi}, B. and {Behroozi}, P. and {Biffi}, V. and {Biviano}, A. and {Lamer}, G. and {Malavasi}, N. and {Mazengo}, D. and {Toptun}, V.},
        title = "{Detecting clusters and groups of galaxies populating the local Universe in large optical spectroscopic surveys}",
      journal = {\aap},
         year = 2025,
        month = feb,
       volume = {694},
          eid = {A207},
        pages = {A207},
          doi = {10.1051/0004-6361/202452028},
archivePrefix = {arXiv},
       eprint = {2411.16455},
 primaryClass = {astro-ph.GA},
       adsurl = {https://ui.adsabs.harvard.edu/abs/2025A&A...694A.207M}
}

@article{Springel2001,
  title = {Populating a cluster of galaxies - I. Results at \fontshape{it}{z}=0},
  volume = {328},
  ISSN = {1365-2966},
  url = {http://dx.doi.org/10.1046/j.1365-8711.2001.04912.x},
  DOI = {10.1046/j.1365-8711.2001.04912.x},
  number = {3},
  journal = {Monthly Notices of the Royal Astronomical Society},
  publisher = {Oxford University Press (OUP)},
  author = {Springel,  Volker and White,  Simon D. M. and Tormen,  Giuseppe and Kauffmann,  Guinevere},
  year = {2001},
  month = dec,
  pages = {726–750}
}

@article{Dolag2009,
  title = {Substructures in hydrodynamical cluster simulations},
  volume = {399},
  ISSN = {1365-2966},
  url = {http://dx.doi.org/10.1111/j.1365-2966.2009.15034.x},
  DOI = {10.1111/j.1365-2966.2009.15034.x},
  number = {2},
  journal = {Monthly Notices of the Royal Astronomical Society},
  publisher = {Oxford University Press (OUP)},
  author = {Dolag,  K. and Borgani,  S. and Murante,  G. and Springel,  V.},
  year = {2009},
  month = oct,
  pages = {497–514}
}

@article{Popesso2025average,
  title = {Average X-ray properties of galaxy groups: From Milky Way-like halos to massive clusters},
  volume = {704},
  ISSN = {1432-0746},
  url = {http://dx.doi.org/10.1051/0004-6361/202453255},
  DOI = {10.1051/0004-6361/202453255},
  journal = {A\&A},
  publisher = {EDP Sciences},
  author = {Popesso,  P. and Marini,  I. and Dolag,  K. and Lamer,  G. and Csizi,  B. and Biffi,  V. and Robothan,  A. and Bravo,  M. and Biviano,  A. and Vladutescu-Zopp,  S. and Lovisari,  L. and Ettori,  S. and Angelinelli,  M. and Driver,  S. and Toptun,  V. and Dev,  A. and Mazengo,  D. and Merloni,  A. and Zhang,  Y. and Comparat,  J. and Ponti,  G. and Mroczkowski,  T. and Bulbul,  E.},
  year = {2025},
  month = dec,
  pages = {A278}
}

@ARTICLE{Jin2025,
       author = {{Jin}, Gaoxiang and {Kauffmann}, Guinevere and {Best}, Philip N. and {Shenoy}, Shravya and {Ma{\l}ek}, Katarzyna},
        title = "{The host galaxies of radio AGN: New views from combining LoTSS and MaNGA observations}",
      journal = {\aap},
         year = 2025,
        month = feb,
       volume = {694},
          eid = {A309},
        pages = {A309},
          doi = {10.1051/0004-6361/202451974},
archivePrefix = {arXiv},
       eprint = {2409.01279},
 primaryClass = {astro-ph.GA},
       adsurl = {https://ui.adsabs.harvard.edu/abs/2025A&A...694A.309J}
}

@ARTICLE{2019MNRAS.485.4817D,
       author = {{Donnari}, Martina and {Pillepich}, Annalisa and {Nelson}, Dylan and {Vogelsberger}, Mark and {Genel}, Shy and {Weinberger}, Rainer and {Marinacci}, Federico and {Springel}, Volker and {Hernquist}, Lars},
        title = "{The star formation activity of IllustrisTNG galaxies: main sequence, UVJ diagram, quenched fractions, and systematics}",
      journal = {\mnras},
         year = 2019,
        month = jun,
       volume = {485},
       number = {4},
        pages = {4817-4840},
          doi = {10.1093/mnras/stz712},
archivePrefix = {arXiv},
       eprint = {1812.07584},
 primaryClass = {astro-ph.GA},
       adsurl = {https://ui.adsabs.harvard.edu/abs/2019MNRAS.485.4817D}
}

@ARTICLE{2019MNRAS.483.3213P,
       author = {{Popesso}, P. and {Concas}, A. and {Morselli}, L. and {Schreiber}, C. and {Rodighiero}, G. and {Cresci}, G. and {Belli}, S. and {Erfanianfar}, G. and {Mancini}, C. and {Inami}, H. and {Dickinson}, M. and {Ilbert}, O. and {Pannella}, M. and {Elbaz}, D.},
        title = "{The main sequence of star-forming galaxies - I. The local relation and its bending}",
      journal = {\mnras},
         year = 2019,
        month = mar,
       volume = {483},
       number = {3},
        pages = {3213-3226},
          doi = {10.1093/mnras/sty3210},
archivePrefix = {arXiv},
       eprint = {1812.07057},
 primaryClass = {astro-ph.GA},
       adsurl = {https://ui.adsabs.harvard.edu/abs/2019MNRAS.483.3213P}
}

@ARTICLE{2019MNRAS.486.2827D,
       author = {{Dav{\'e}}, Romeel and {Angl{\'e}s-Alc{\'a}zar}, Daniel and {Narayanan}, Desika and {Li}, Qi and {Rafieferantsoa}, Mika H. and {Appleby}, Sarah},
        title = "{SIMBA: Cosmological simulations with black hole growth and feedback}",
      journal = {\mnras},
         year = 2019,
        month = jun,
       volume = {486},
       number = {2},
        pages = {2827-2849},
          doi = {10.1093/mnras/stz937},
archivePrefix = {arXiv},
       eprint = {1901.10203},
 primaryClass = {astro-ph.GA},
       adsurl = {https://ui.adsabs.harvard.edu/abs/2019MNRAS.486.2827D}
}

@ARTICLE{2023MNRAS.519.1526P,
       author = {{Popesso}, P. and {Concas}, A. and {Cresci}, G. and {Belli}, S. and {Rodighiero}, G. and {Inami}, H. and {Dickinson}, M. and {Ilbert}, O. and {Pannella}, M. and {Elbaz}, D.},
        title = "{The main sequence of star-forming galaxies across cosmic times}",
      journal = {\mnras},
         year = 2023,
        month = feb,
       volume = {519},
       number = {1},
        pages = {1526-1544},
          doi = {10.1093/mnras/stac3214},
archivePrefix = {arXiv},
       eprint = {2203.10487},
 primaryClass = {astro-ph.GA},
       adsurl = {https://ui.adsabs.harvard.edu/abs/2023MNRAS.519.1526P}
}

@ARTICLE{2025MNRAS.538..976M,
       author = {{Mun}, Marcie and {Wisnioski}, Emily and {Harborne}, Katherine E. and {Lagos}, Claudia D.~P. and {Valenzuela}, Lucas M. and {Remus}, Rhea-Silvia and {Mendel}, J. Trevor and {Battisti}, Andrew J. and {Ellison}, Sara L. and {Foster}, Caroline and {Bravo}, Matias and {Brough}, Sarah and {Croom}, Scott M. and {Gao}, Tianmu and {Grasha}, Kathryn and {Gupta}, Anshu and {Mai}, Yifan and {Mailvaganam}, Anilkumar and {Muller}, Eric G.~M. and {Sharma}, Gauri and {Sweet}, Sarah M. and {Taylor}, Edward N. and {Zafar}, Tayyaba},
        title = "{The MAGPI Survey: radial trends in star formation across different cosmological simulations in comparison with observations at z \raisebox{-0.5ex}\textasciitilde 0.3}",
      journal = {\mnras},
         year = 2025,
        month = apr,
       volume = {538},
       number = {2},
        pages = {976-997},
          doi = {10.1093/mnras/staf342},
archivePrefix = {arXiv},
       eprint = {2411.17882},
 primaryClass = {astro-ph.GA},
       adsurl = {https://ui.adsabs.harvard.edu/abs/2025MNRAS.538..976M}
}

@ARTICLE{2020MNRAS.499..230B,
       author = {{Bluck}, Asa F.~L. and {Maiolino}, Roberto and {Piotrowska}, Joanna M. and {Trussler}, James and {Ellison}, Sara L. and {S{\'a}nchez}, Sebastian F. and {Thorp}, Mallory D. and {Teimoorinia}, Hossen and {Moreno}, Jorge and {Conselice}, Christopher J.},
        title = "{How do central and satellite galaxies quench? - Insights from spatially resolved spectroscopy in the MaNGA survey}",
      journal = {\mnras},
         year = 2020,
        month = nov,
       volume = {499},
       number = {1},
        pages = {230-268},
          doi = {10.1093/mnras/staa2806},
archivePrefix = {arXiv},
       eprint = {2009.05341},
 primaryClass = {astro-ph.GA},
       adsurl = {https://ui.adsabs.harvard.edu/abs/2020MNRAS.499..230B}
}

@ARTICLE{2010ApJ...721..193P,
       author = {{Peng}, Ying-jie and {Lilly}, Simon J. and {Kova{\v{c}}}, Katarina and {Bolzonella}, Micol and {Pozzetti}, Lucia and {Renzini}, Alvio and {Zamorani}, Gianni and {Ilbert}, Olivier and {Knobel}, Christian and {Iovino}, Angela and {Maier}, Christian and {Cucciati}, Olga and {Tasca}, Lidia and {Carollo}, C. Marcella and {Silverman}, John and {Kampczyk}, Pawel and {de Ravel}, Loic and {Sanders}, David and {Scoville}, Nicholas and {Contini}, Thierry and {Mainieri}, Vincenzo and {Scodeggio}, Marco and {Kneib}, Jean-Paul and {Le F{\`e}vre}, Olivier and {Bardelli}, Sandro and {Bongiorno}, Angela and {Caputi}, Karina and {Coppa}, Graziano and {de la Torre}, Sylvain and {Franzetti}, Paolo and {Garilli}, Bianca and {Lamareille}, Fabrice and {Le Borgne}, Jean-Francois and {Le Brun}, Vincent and {Mignoli}, Marco and {Perez Montero}, Enrique and {Pello}, Roser and {Ricciardelli}, Elena and {Tanaka}, Masayuki and {Tresse}, Laurence and {Vergani}, Daniela and {Welikala}, Niraj and {Zucca}, Elena and {Oesch}, Pascal and {Abbas}, Ummi and {Barnes}, Luke and {Bordoloi}, Rongmon and {Bottini}, Dario and {Cappi}, Alberto and {Cassata}, Paolo and {Cimatti}, Andrea and {Fumana}, Marco and {Hasinger}, Gunther and {Koekemoer}, Anton and {Leauthaud}, Alexei and {Maccagni}, Dario and {Marinoni}, Christian and {McCracken}, Henry and {Memeo}, Pierdomenico and {Meneux}, Baptiste and {Nair}, Preethi and {Porciani}, Cristiano and {Presotto}, Valentina and {Scaramella}, Roberto},
        title = "{Mass and Environment as Drivers of Galaxy Evolution in SDSS and zCOSMOS and the Origin of the Schechter Function}",
      journal = {\apj},
         year = 2010,
        month = sep,
       volume = {721},
       number = {1},
        pages = {193-221},
          doi = {10.1088/0004-637X/721/1/193},
archivePrefix = {arXiv},
       eprint = {1003.4747},
 primaryClass = {astro-ph.CO},
       adsurl = {https://ui.adsabs.harvard.edu/abs/2010ApJ...721..193P}
}

@ARTICLE{2026A&A...706A.376N,
       author = {{Nyiransengiyumva}, Beatrice and {Povi{\'c}}, Mirjana and {Nkundabakura}, Pheneas and {Mutabazi}, Tom and {Mahoro}, Antoine},
        title = "{The impact of selection criteria on the properties of green valley galaxies}",
      journal = {\aap},
         year = 2026,
        month = feb,
       volume = {706},
          eid = {A376},
        pages = {A376},
          doi = {10.1051/0004-6361/202555290},
archivePrefix = {arXiv},
       eprint = {2512.20379},
 primaryClass = {astro-ph.GA},
       adsurl = {https://ui.adsabs.harvard.edu/abs/2026A&A...706A.376N}
}

@ARTICLE{2015ApJ...801L..29R,
       author = {{Renzini}, Alvio and {Peng}, Ying-jie},
        title = "{An Objective Definition for the Main Sequence of Star-forming Galaxies}",
      journal = {\apjl},
         year = 2015,
        month = mar,
       volume = {801},
       number = {2},
          eid = {L29},
        pages = {L29},
          doi = {10.1088/2041-8205/801/2/L29},
archivePrefix = {arXiv},
       eprint = {1502.01027},
 primaryClass = {astro-ph.GA},
       adsurl = {https://ui.adsabs.harvard.edu/abs/2015ApJ...801L..29R}
}

@ARTICLE{2019MNRAS.490.5285P,
       author = {{Popesso}, P. and {Morselli}, L. and {Concas}, A. and {Schreiber}, C. and {Rodighiero}, G. and {Cresci}, G. and {Belli}, S. and {Ilbert}, O. and {Erfanianfar}, G. and {Mancini}, C. and {Inami}, H. and {Dickinson}, M. and {Pannella}, M. and {Elbaz}, D.},
        title = "{The main sequence of star-forming galaxies - II. A non-evolving slope at the high-mass end}",
      journal = {\mnras},
         year = 2019,
        month = dec,
       volume = {490},
       number = {4},
        pages = {5285-5299},
          doi = {10.1093/mnras/stz2635},
archivePrefix = {arXiv},
       eprint = {1909.07760},
 primaryClass = {astro-ph.GA},
       adsurl = {https://ui.adsabs.harvard.edu/abs/2019MNRAS.490.5285P}
}

@ARTICLE{2018MNRAS.477.3014B,
       author = {{Belfiore}, Francesco and {Maiolino}, Roberto and {Bundy}, Kevin and {Masters}, Karen and {Bershady}, Matthew and {Oyarz{\'u}n}, Grecco A. and {Lin}, Lihwai and {Cano-Diaz}, Mariana and {Wake}, David and {Spindler}, Ashley and {Thomas}, Daniel and {Brownstein}, Joel R. and {Drory}, Niv and {Yan}, Renbin},
        title = "{SDSS IV MaNGA - sSFR profiles and the slow quenching of discs in green valley galaxies}",
      journal = {\mnras},
         year = 2018,
        month = jul,
       volume = {477},
       number = {3},
        pages = {3014-3029},
          doi = {10.1093/mnras/sty768},
archivePrefix = {arXiv},
       eprint = {1710.05034},
 primaryClass = {astro-ph.GA},
       adsurl = {https://ui.adsabs.harvard.edu/abs/2018MNRAS.477.3014B}
}

@ARTICLE{2024A&A...690A.206V,
       author = {{Valenzuela}, Lucas M. and {Remus}, Rhea-Silvia and {Dolag}, Klaus and {Seidel}, Benjamin A.},
        title = "{Galaxy shapes in Magneticum: I. Connecting stellar and dark matter shapes to dynamical and morphological galaxy properties and the large-scale structure}",
      journal = {\aap},
         year = 2024,
        month = oct,
       volume = {690},
          eid = {A206},
        pages = {A206},
          doi = {10.1051/0004-6361/202450184},
archivePrefix = {arXiv},
       eprint = {2404.01368},
 primaryClass = {astro-ph.GA},
       adsurl = {https://ui.adsabs.harvard.edu/abs/2024A&A...690A.206V}
}

@ARTICLE{2024MNRAS.534..361S,
       author = {{Scharr{\'e}}, Lucie and {Sorini}, Daniele and {Dav{\'e}}, Romeel},
        title = "{The effects of stellar and AGN feedback on the cosmic star formation history in the SIMBA simulations}",
      journal = {\mnras},
         year = 2024,
        month = oct,
       volume = {534},
       number = {1},
        pages = {361-383},
          doi = {10.1093/mnras/stae2098},
archivePrefix = {arXiv},
       eprint = {2404.07252},
 primaryClass = {astro-ph.GA},
       adsurl = {https://ui.adsabs.harvard.edu/abs/2024MNRAS.534..361S}
}

@ARTICLE{2004MNRAS.353..713K,
       author = {{Kauffmann}, Guinevere and {White}, Simon D.~M. and {Heckman}, Timothy M. and {M{\'e}nard}, Brice and {Brinchmann}, Jarle and {Charlot}, St{\'e}phane and {Tremonti}, Christy and {Brinkmann}, Jon},
        title = "{The environmental dependence of the relations between stellar mass, structure, star formation and nuclear activity in galaxies}",
      journal = {\mnras},
         year = 2004,
        month = sep,
       volume = {353},
       number = {3},
        pages = {713-731},
          doi = {10.1111/j.1365-2966.2004.08117.x},
archivePrefix = {arXiv},
       eprint = {astro-ph/0402030},
 primaryClass = {astro-ph},
       adsurl = {https://ui.adsabs.harvard.edu/abs/2004MNRAS.353..713K}
}

@ARTICLE{2019MNRAS.488.3143B,
       author = {{Behroozi}, Peter and {Wechsler}, Risa H. and {Hearin}, Andrew P. and {Conroy}, Charlie},
        title = "{UNIVERSEMACHINE: The correlation between galaxy growth and dark matter halo assembly from z = 0-10}",
      journal = {\mnras},
         year = 2019,
        month = sep,
       volume = {488},
       number = {3},
        pages = {3143-3194},
          doi = {10.1093/mnras/stz1182},
archivePrefix = {arXiv},
       eprint = {1806.07893},
 primaryClass = {astro-ph.GA},
       adsurl = {https://ui.adsabs.harvard.edu/abs/2019MNRAS.488.3143B}
}

@ARTICLE{2015AJ....150...19L,
       author = {{Law}, David R. and {Yan}, Renbin and {Bershady}, Matthew A. and {Bundy}, Kevin and {Cherinka}, Brian and {Drory}, Niv and {MacDonald}, Nicholas and {S{\'a}nchez-Gallego}, Jos{\'e} R. and {Wake}, David A. and {Weijmans}, Anne-Marie and {Blanton}, Michael R. and {Klaene}, Mark A. and {Moran}, Sean M. and {Sanchez}, Sebastian F. and {Zhang}, Kai},
        title = "{Observing Strategy for the SDSS-IV/MaNGA IFU Galaxy Survey}",
      journal = {\aj},
         year = 2015,
        month = jul,
       volume = {150},
       number = {1},
          eid = {19},
        pages = {19},
          doi = {10.1088/0004-6256/150/1/19},
archivePrefix = {arXiv},
       eprint = {1505.04285},
 primaryClass = {astro-ph.IM},
       adsurl = {https://ui.adsabs.harvard.edu/abs/2015AJ....150...19L}
}

@ARTICLE{2015ApJ...798....7B,
       author = {{Bundy}, Kevin and {Bershady}, Matthew A. and {Law}, David R. and {Yan}, Renbin and {Drory}, Niv and {MacDonald}, Nicholas and {Wake}, David A. and {Cherinka}, Brian and {S{\'a}nchez-Gallego}, Jos{\'e} R. and {Weijmans}, Anne-Marie and {Thomas}, Daniel and {Tremonti}, Christy and {Masters}, Karen and {Coccato}, Lodovico and {Diamond-Stanic},
       Aleksandar M. and {Arag{\'o}n-Salamanca}, Alfonso and {Avila-Reese}, Vladimir and {Badenes}, Carles and {Falc{\'o}n-Barroso}, J{\'e}sus and {Belfiore}, Francesco and {Bizyaev}, Dmitry and {Blanc}, Guillermo A. and {Bland-Hawthorn}, Joss and {Blanton}, Michael R. and {Brownstein}, Joel R. and {Byler}, Nell and {Cappellari}, Michele and {Conroy}, Charlie and {Dutton}, Aaron A. and {Emsellem}, Eric and {Etherington}, James and {Frinchaboy}, Peter M. and {Fu}, Hai and {Gunn}, James E. and {Harding}, Paul and {Johnston}, Evelyn J. and {Kauffmann}, Guinevere and {Kinemuchi}, Karen and {Klaene}, Mark A. and {Knapen}, Johan H. and {Leauthaud}, Alexie and {Li}, Cheng and {Lin}, Lihwai and {Maiolino}, Roberto and {Malanushenko}, Viktor and {Malanushenko}, Elena and {Mao}, Shude and {Maraston}, Claudia and {McDermid}, Richard M. and {Merrifield}, Michael R. and {Nichol}, Robert C. and {Oravetz}, Daniel and {Pan}, Kaike and {Parejko}, John K. and {Sanchez}, Sebastian F. and {Schlegel}, David and {Simmons}, Audrey and {Steele}, Oliver and {Steinmetz}, Matthias and {Thanjavur}, Karun and {Thompson}, Benjamin A. and {Tinker}, Jeremy L. and {van den Bosch}, Remco C.~E. and {Westfall}, Kyle B. and {Wilkinson}, David and {Wright}, Shelley and {Xiao}, Ting and {Zhang}, Kai},
        title = "{Overview of the SDSS-IV MaNGA Survey: Mapping nearby Galaxies at Apache Point Observatory}",
      journal = {\apj},
         year = 2015,
        month = jan,
       volume = {798},
       number = {1},
          eid = {7},
        pages = {7},
          doi = {10.1088/0004-637X/798/1/7},
archivePrefix = {arXiv},
       eprint = {1412.1482},
 primaryClass = {astro-ph.GA},
       adsurl = {https://ui.adsabs.harvard.edu/abs/2015ApJ...798....7B}
}

@ARTICLE{2022ApJS..259...35A,
       author = {{Abdurro'uf} and {Accetta}, Katherine and {Aerts}, Conny and {Silva Aguirre}, V{\'\i}ctor and {Ahumada}, Romina and {Ajgaonkar}, Nikhil and {Filiz Ak}, N. and {Alam}, Shadab and {Allende Prieto}, Carlos and {Almeida}, Andr{\'e}s and {Anders}, Friedrich and {Anderson}, Scott F. and {Andrews}, Brett H. and {Anguiano}, Borja and {Aquino-Ort{\'\i}z}, Erik and {Arag{\'o}n-Salamanca}, Alfonso and {Argudo-Fern{\'a}ndez}, Maria and {Ata}, Metin and {Aubert}, Marie and {Avila-Reese}, Vladimir and {Badenes}, Carles and {Barb{\'a}}, Rodolfo H. and {Barger}, Kat and {Barrera-Ballesteros}, Jorge K. and {Beaton}, Rachael L. and {Beers}, Timothy C. and {Belfiore}, Francesco and {Bender}, Chad F. and {Bernardi}, Mariangela and {Bershady}, Matthew A. and {Beutler}, Florian and {Bidin}, Christian Moni and {Bird}, Jonathan C. and {Bizyaev}, Dmitry and {Blanc}, Guillermo A. and {Blanton}, Michael R. and {Boardman}, Nicholas Fraser and {Bolton}, Adam S. and {Boquien}, M{\'e}d{\'e}ric and {Borissova}, Jura and {Bovy}, Jo and {Brandt}, W.~N. and {Brown}, Jordan and {Brownstein}, Joel R. and {Brusa}, Marcella and {Buchner}, Johannes and {Bundy}, Kevin and {Burchett}, Joseph N. and {Bureau}, Martin and {Burgasser}, Adam and {Cabang}, Tuesday K. and {Campbell}, Stephanie and {Cappellari}, Michele and {Carlberg}, Joleen K. and {Wanderley}, F{\'a}bio Carneiro and {Carrera}, Ricardo and {Cash}, Jennifer and {Chen}, Yan-Ping and {Chen}, Wei-Huai and {Cherinka}, Brian and {Chiappini}, Cristina and {Choi}, Peter Doohyun and {Chojnowski}, S. Drew and {Chung}, Haeun and {Clerc}, Nicolas and {Cohen}, Roger E. and {Comerford}, Julia M. and {Comparat}, Johan and {da Costa}, Luiz and {Covey}, Kevin and {Crane}, Jeffrey D. and {Cruz-Gonzalez}, Irene and {Culhane}, Connor and {Cunha}, Katia and {Dai}, Y. Sophia and {Damke}, Guillermo and {Darling}, Jeremy and {Davidson}, Jr., James W. and {Davies}, Roger and {Dawson}, Kyle and {De Lee}, Nathan and {Diamond-Stanic}, Aleksandar M. and {Cano-D{\'\i}az}, Mariana and {S{\'a}nchez}, Helena Dom{\'\i}nguez and {Donor}, John and {Duckworth}, Chris and {Dwelly}, Tom and {Eisenstein}, Daniel J. and {Elsworth}, Yvonne P. and {Emsellem}, Eric and {Eracleous}, Mike and {Escoffier}, Stephanie and {Fan}, Xiaohui and {Farr}, Emily and {Feng}, Shuai and {Fern{\'a}ndez-Trincado}, Jos{\'e} G. and {Feuillet}, Diane and {Filipp}, Andreas and {Fillingham}, Sean P. and {Frinchaboy}, Peter M. and {Fromenteau}, Sebastien and {Galbany}, Llu{\'\i}s and {Garc{\'\i}a}, Rafael A. and {Garc{\'\i}a-Hern{\'a}ndez}, D.~A. and {Ge}, Junqiang and {Geisler}, Doug and {Gelfand}, Joseph and {G{\'e}ron}, Tobias and {Gibson}, Benjamin J. and {Goddy}, Julian and {Godoy-Rivera}, Diego and {Grabowski}, Kathleen and {Green}, Paul J. and {Greener}, Michael and {Grier}, Catherine J. and {Griffith}, Emily and {Guo}, Hong and {Guy}, Julien and {Hadjara}, Massinissa and {Harding}, Paul and {Hasselquist}, Sten and {Hayes}, Christian R. and {Hearty}, Fred and {Hern{\'a}ndez}, Jes{\'u}s and {Hill}, Lewis and {Hogg}, David W. and {Holtzman}, Jon A. and {Horta}, Danny and {Hsieh}, Bau-Ching and {Hsu}, Chin-Hao and {Hsu}, Yun-Hsin and {Huber}, Daniel and {Huertas-Company}, Marc and {Hutchinson}, Brian and {Hwang}, Ho Seong and {Ibarra-Medel}, H{\'e}ctor J. and {Chitham}, Jacob Ider and {Ilha}, Gabriele S. and {Imig}, Julie and {Jaekle}, Will and {Jayasinghe}, Tharindu and {Ji}, Xihan and {Johnson}, Jennifer A. and {Jones}, Amy and {J{\"o}nsson}, Henrik and {Katkov}, Ivan and {Khalatyan}, Dr., Arman and {Kinemuchi}, Karen and {Kisku}, Shobhit and {Knapen}, Johan H. and {Kneib}, Jean-Paul and {Kollmeier}, Juna A. and {Kong}, Miranda and {Kounkel}, Marina and {Kreckel}, Kathryn and {Krishnarao}, Dhanesh and {Lacerna}, Ivan and {Lane}, Richard R. and {Langgin}, Rachel and {Lavender}, Ramon and {Law}, David R. and {Lazarz}, Daniel and {Leung}, Henry W. and {Leung}, Ho-Hin and {Lewis}, Hannah M. and {Li}, Cheng and {Li}, Ran and {Lian}, Jianhui and {Liang}, Fu-Heng and {Lin}, Lihwai and {Lin}, Yen-Ting and {Lin}, Sicheng and {Lintott}, Chris and {Long}, Dan and {Longa-Pe{\~n}a}, Pen{\'e}lope and {L{\'o}pez-Cob{\'a}}, Carlos and {Lu}, Shengdong and {Lundgren}, Britt F. and {Luo}, Yuanze and {Mackereth}, J. Ted and {de la Macorra}, Axel and {Mahadevan}, Suvrath and {Majewski}, Steven R. and {Manchado}, Arturo and {Mandeville}, Travis and {Maraston}, Claudia and {Margalef-Bentabol}, Berta and {Masseron}, Thomas and {Masters}, Karen L. and {Mathur}, Savita and {McDermid}, Richard M. and {Mckay}, Myles and {Merloni}, Andrea and {Merrifield}, Michael and {Meszaros}, Szabolcs and {Miglio}, Andrea and {Di Mille}, Francesco and {Minniti}, Dante and {Minsley}, Rebecca and {Monachesi}, Antonela},
        title = "{The Seventeenth Data Release of the Sloan Digital Sky Surveys: Complete Release of MaNGA, MaStar, and APOGEE-2 Data}",
      journal = {\apjs},
         year = 2022,
        month = apr,
       volume = {259},
       number = {2},
          eid = {35},
        pages = {35},
          doi = {10.3847/1538-4365/ac4414},
archivePrefix = {arXiv},
       eprint = {2112.02026},
 primaryClass = {astro-ph.GA},
       adsurl = {https://ui.adsabs.harvard.edu/abs/2022ApJS..259...35A}
}

@ARTICLE{2022ApJS..262...36S,
       author = {{S{\'a}nchez}, S.~F. and {Barrera-Ballesteros}, J.~K. and {Lacerda}, E. and {Mej{\'\i}a-Narvaez}, A. and {Camps-Fari{\~n}a}, A. and {Bruzual}, Gustavo and {Espinosa-Ponce}, C. and {Rodr{\'\i}guez-Puebla}, A. and {Calette}, A.~R. and {Ibarra-Medel}, H. and {Avila-Reese}, V. and {Hernandez-Toledo}, H. and {Bershady}, M.~A. and {Cano-Diaz}, M. and {Munguia-Cordova}, A.~M.},
        title = "{SDSS-IV MaNGA: pyPipe3D Analysis Release for 10,000 Galaxies}",
      journal = {\apjs},
         year = 2022,
        month = oct,
       volume = {262},
       number = {2},
          eid = {36},
        pages = {36},
          doi = {10.3847/1538-4365/ac7b8f},
archivePrefix = {arXiv},
       eprint = {2206.07062},
 primaryClass = {astro-ph.GA},
       adsurl = {https://ui.adsabs.harvard.edu/abs/2022ApJS..262...36S}
}

@ARTICLE{2013MNRAS.435.3444P,
       author = {{Povi{\'c}}, M. and {Huertas-Company}, M. and {Aguerri}, J.~A.~L. and {M{\'a}rquez}, I. and {Masegosa}, J. and {Husillos}, C. and {Molino}, A. and {Crist{\'o}bal-Hornillos}, D. and {Perea}, J. and {Ben{\'\i}tez}, N. and {Olmo}, A. del and {Fern{\'a}ndez-Soto}, A. and {Jim{\'e}nez-Teja}, Y. and {Moles}, M. and {Alfaro}, E. and {Aparicio-Villegas}, T. and {Ascaso}, B. and {Broadhurst}, T. and {Cabrera-Ca{\~n}o}, J. and {Castander}, F.~J. and {Cepa}, J. and {Fernandez Lorenzo}, M. and {Cervi{\~n}o}, M. and {Delgado}, R.~M. Gonz{\'a}lez and {Infante}, L. and {L{\'o}pez-Sanjuan}, C. and {Mart{\'\i}nez}, V.~J. and {Matute}, I. and {Oteo}, I. and {P{\'e}rez-Garc{\'\i}a}, A.~M. and {Prada}, F. and {Quintana}, J.~M.},
        title = "{The ALHAMBRA survey: reliable morphological catalogue of 22 051 early- and late-type galaxies}",
      journal = {\mnras},
         year = 2013,
        month = nov,
       volume = {435},
       number = {4},
        pages = {3444-3461},
          doi = {10.1093/mnras/stt1538},
archivePrefix = {arXiv},
       eprint = {1308.3146},
 primaryClass = {astro-ph.CO},
       adsurl = {https://ui.adsabs.harvard.edu/abs/2013MNRAS.435.3444P}
}

@ARTICLE{2021MNRAS.500.2036K,
       author = {{Katsianis}, Antonios and {Xu}, Haojie and {Yang}, Xiaohu and {Luo}, Yu and {Cui}, Weiguang and {Dav{\'e}}, Romeel and {Lagos}, Claudia Del P. and {Zheng}, Xianzhong and {Zhao}, Ping},
        title = "{The specific star formation rate function at different mass scales and quenching: a comparison between cosmological models and SDSS}",
      journal = {\mnras},
         year = 2021,
        month = jan,
       volume = {500},
       number = {2},
        pages = {2036-2048},
          doi = {10.1093/mnras/staa3236},
archivePrefix = {arXiv},
       eprint = {2010.08173},
 primaryClass = {astro-ph.GA},
       adsurl = {https://ui.adsabs.harvard.edu/abs/2021MNRAS.500.2036K}
}

@ARTICLE{2024A&A...683A..57R,
       author = {{Rihtar{\v{s}}i{\v{c}}}, G. and {Biffi}, V. and {Fabjan}, D. and {Dolag}, K.},
        title = "{Environmental dependence of AGN activity and star formation in galaxy clusters from Magneticum simulations}",
      journal = {\aap},
         year = 2024,
        month = mar,
       volume = {683},
          eid = {A57},
        pages = {A57},
          doi = {10.1051/0004-6361/202347444},
archivePrefix = {arXiv},
       eprint = {2307.06374},
 primaryClass = {astro-ph.GA},
       adsurl = {https://ui.adsabs.harvard.edu/abs/2024A&A...683A..57R}
}

@ARTICLE{2016MNRAS.458L..34E,
       author = {{Ellison}, Sara L. and {Teimoorinia}, Hossen and {Rosario}, David J. and {Mendel}, J. Trevor},
        title = "{The star formation rates of active galactic nuclei host galaxies}",
      journal = {\mnras},
         year = 2016,
        month = may,
       volume = {458},
       number = {1},
        pages = {L34-L38},
          doi = {10.1093/mnrasl/slw012},
archivePrefix = {arXiv},
       eprint = {1601.03349},
 primaryClass = {astro-ph.GA},
       adsurl = {https://ui.adsabs.harvard.edu/abs/2016MNRAS.458L..34E}
}

@ARTICLE{Comerford_2020,
       author = {{Comerford}, Julia M. and {Negus}, James and {M{\"u}ller-S{\'a}nchez}, Francisco and {Eracleous}, Michael and {Wylezalek}, Dominika and {Storchi-Bergmann}, Thaisa and {Greene}, Jenny E. and {Barrows}, R. Scott and {Nevin}, Rebecca and {Roy}, Namrata and {Stemo}, Aaron},
        title = "{A Catalog of 406 AGNs in MaNGA: A Connection between Radio-mode AGNs and Star Formation Quenching}",
      journal = {\apj},
         year = 2020,
        month = oct,
       volume = {901},
       number = {2},
          eid = {159},
        pages = {159},
          doi = {10.3847/1538-4357/abb2ae},
archivePrefix = {arXiv},
       eprint = {2008.11210},
 primaryClass = {astro-ph.GA},
       adsurl = {https://ui.adsabs.harvard.edu/abs/2020ApJ...901..159C}
}

@ARTICLE{2013ApJ...770...57B,
       author = {{Behroozi}, Peter S. and {Wechsler}, Risa H. and {Conroy}, Charlie},
        title = "{The Average Star Formation Histories of Galaxies in Dark Matter Halos from z = 0-8}",
      journal = {\apj},
         year = 2013,
        month = jun,
       volume = {770},
       number = {1},
          eid = {57},
        pages = {57},
          doi = {10.1088/0004-637X/770/1/57},
archivePrefix = {arXiv},
       eprint = {1207.6105},
 primaryClass = {astro-ph.CO},
       adsurl = {https://ui.adsabs.harvard.edu/abs/2013ApJ...770...57B}
}

@ARTICLE{2025A&A...700A.167T,
       author = {{Toptun}, V. and {Popesso}, P. and {Marini}, I. and {Dolag}, K. and {Lamer}, G. and {Yang}, X. and {Li}, Q. and {Csizi}, B. and {Lovisari}, L. and {Ettori}, S. and {Biffi}, V. and {Vladutescu-Zopp}, S. and {Dev}, A. and {Mazengo}, D. and {Merloni}, A. and {Comparat}, J. and {Ponti}, G. and {Bulbul}, E.},
        title = "{The eROSITA view on the halo mass{\textendash}temperature relation: From low-mass groups to massive clusters}",
      journal = {\aap},
         year = 2025,
        month = aug,
       volume = {700},
          eid = {A167},
        pages = {A167},
          doi = {10.1051/0004-6361/202554352},
archivePrefix = {arXiv},
       eprint = {2505.01502},
 primaryClass = {astro-ph.GA},
       adsurl = {https://ui.adsabs.harvard.edu/abs/2025A&A...700A.167T}
}

@ARTICLE{2023ApJ...952...12M,
       author = {{Mahoro}, Antoine and {V{\"a}is{\"a}nen}, Petri and {Povi{\'c}}, Mirjana and {Nkundabakura}, Pheneas and {van der Heyden}, Kurt and {Cazzoli}, Sara and {Worku}, Samuel B. and {M{\'a}rquez}, Isabel and {Masegosa}, Josefa and {Randriamampandry}, Solohery M. and {Mogotsi}, Moses},
        title = "{The [O III] Profiles of Far-infrared Active and Inactive Optically Selected Green Valley Galaxies}",
      journal = {\apj},
         year = 2023,
        month = jul,
       volume = {952},
       number = {1},
          eid = {12},
        pages = {12},
          doi = {10.3847/1538-4357/accea1},
archivePrefix = {arXiv},
       eprint = {2304.09284},
 primaryClass = {astro-ph.GA},
       adsurl = {https://ui.adsabs.harvard.edu/abs/2023ApJ...952...12M}
}

@INPROCEEDINGS{2015IAUS..309..145R,
       author = {{Remus}, Rhea-Silvia and {Dolag}, Klaus and {Bachmann}, Lisa K. and {Beck}, Alexander M. and {Burkert}, Andreas and {Hirschmann}, Michaela and {Teklu}, Adelheid},
        title = "{Disk Galaxies in the Magneticum Pathfinder Simulations}",
    booktitle = {Galaxies in 3D across the Universe},
         year = 2015,
       editor = {{Ziegler}, Bodo L. and {Combes}, Fran{\c{c}}oise and {Dannerbauer}, Helmut and {Verdugo}, Miguel},
       series = {IAU Symposium},
       volume = {309},
        month = feb,
        pages = {145-148},
          doi = {10.1017/S1743921314009491},
       adsurl = {https://ui.adsabs.harvard.edu/abs/2015IAUS..309..145R}
}

@INPROCEEDINGS{2019trec.confE..26D,
       author = {{Dolag}, Klaus},
        title = "{Galaxy Clusters in the Magneticum Simulations}",
    booktitle = {Tracing Cosmic Evolution with Clusters of Galaxies},
         year = 2019,
        month = jul,
          eid = {26},
        pages = {26},
       adsurl = {https://ui.adsabs.harvard.edu/abs/2019trec.confE..26D}
}

@ARTICLE{2022A&A...661A..17B,
       author = {{Biffi}, Veronica and {Dolag}, Klaus and {Reiprich}, Thomas H. and {Veronica}, Angie and {Ramos-Ceja}, Miriam E. and {Bulbul}, Esra and {Ota}, Naomi and {Ghirardini}, Vittorio},
        title = "{The eROSITA view of the Abell 3391/95 field: Case study from the Magneticum cosmological simulation}",
      journal = {\aap},
         year = 2022,
        month = may,
       volume = {661},
          eid = {A17},
        pages = {A17},
          doi = {10.1051/0004-6361/202141107},
archivePrefix = {arXiv},
       eprint = {2106.14542},
 primaryClass = {astro-ph.CO},
       adsurl = {https://ui.adsabs.harvard.edu/abs/2022A&A...661A..17B}
}

@ARTICLE{2015MNRAS.451.4277D,
       author = {{Dolag}, K. and {Gaensler}, B.~M. and {Beck}, A.~M. and {Beck}, M.~C.},
        title = "{Constraints on the distribution and energetics of fast radio bursts using cosmological hydrodynamic simulations}",
      journal = {\mnras},
         year = 2015,
        month = aug,
       volume = {451},
       number = {4},
        pages = {4277-4289},
          doi = {10.1093/mnras/stv1190},
archivePrefix = {arXiv},
       eprint = {1412.4829},
 primaryClass = {astro-ph.CO},
       adsurl = {https://ui.adsabs.harvard.edu/abs/2015MNRAS.451.4277D}
}

@ARTICLE{2024ApJ...971...69D,
       author = {{de Is{\'\i}dio}, Natanael G. and {Men{\'e}ndez-Delmestre}, K. and {Gon{\c{c}}alves}, T.~S. and {Grossi}, M. and {Rodrigues}, D.~C. and {Garavito-Camargo}, N. and {Araujo-Carvalho}, A. and {Beaklini}, P.~P.~B. and {Cavalcante-Coelho}, Y. and {Cortesi}, A. and {Quiroga-Nu{\~n}ez}, L.~H. and {Randriamampandry}, T.},
        title = "{Dark Matter Distribution in Milky Way analog Galaxies}",
      journal = {\apj},
         year = 2024,
        month = aug,
       volume = {971},
       number = {1},
          eid = {69},
        pages = {69},
          doi = {10.3847/1538-4357/ad53c8},
archivePrefix = {arXiv},
       eprint = {2310.13839},
 primaryClass = {astro-ph.GA},
       adsurl = {https://ui.adsabs.harvard.edu/abs/2024ApJ...971...69D}
}

@ARTICLE{2010ApJ...710..903M,
       author = {{Moster}, Benjamin P. and {Somerville}, Rachel S. and {Maulbetsch}, Christian and {van den Bosch}, Frank C. and {Macci{\`o}}, Andrea V. and {Naab}, Thorsten and {Oser}, Ludwig},
        title = "{Constraints on the Relationship between Stellar Mass and Halo Mass at Low and High Redshift}",
      journal = {\apj},
         year = 2010,
        month = feb,
       volume = {710},
       number = {2},
        pages = {903-923},
          doi = {10.1088/0004-637X/710/2/903},
archivePrefix = {arXiv},
       eprint = {0903.4682},
 primaryClass = {astro-ph.CO},
       adsurl = {https://ui.adsabs.harvard.edu/abs/2010ApJ...710..903M}
}

@ARTICLE{2022ApJ...933..161M,
       author = {{McDonough}, Bryanne and {Brainerd}, Tereasa G.},
        title = "{The Distribution of Satellite Galaxies in the IllustrisTNG100 Simulation}",
      journal = {\apj},
         year = 2022,
        month = jul,
       volume = {933},
       number = {2},
          eid = {161},
        pages = {161},
          doi = {10.3847/1538-4357/ac752d},
archivePrefix = {arXiv},
       eprint = {2206.00045},
 primaryClass = {astro-ph.GA},
       adsurl = {https://ui.adsabs.harvard.edu/abs/2022ApJ...933..161M}
}

@ARTICLE{2018MNRAS.475..624N,
       author = {{Nelson}, Dylan and {Pillepich}, Annalisa and {Springel}, Volker and {Weinberger}, Rainer and {Hernquist}, Lars and {Pakmor}, R{\"u}diger and {Genel}, Shy and {Torrey}, Paul and {Vogelsberger}, Mark and {Kauffmann}, Guinevere and {Marinacci}, Federico and {Naiman}, Jill},
        title = "{First results from the IllustrisTNG simulations: the galaxy colour bimodality}",
      journal = {\mnras},
         year = 2018,
        month = mar,
       volume = {475},
       number = {1},
        pages = {624-647},
          doi = {10.1093/mnras/stx3040},
archivePrefix = {arXiv},
       eprint = {1707.03395},
 primaryClass = {astro-ph.GA},
       adsurl = {https://ui.adsabs.harvard.edu/abs/2018MNRAS.475..624N}
}

@ARTICLE{2018MNRAS.475..676S,
       author = {{Springel}, Volker and {Pakmor}, R{\"u}diger and {Pillepich}, Annalisa and {Weinberger}, Rainer and {Nelson}, Dylan and {Hernquist}, Lars and {Vogelsberger}, Mark and {Genel}, Shy and {Torrey}, Paul and {Marinacci}, Federico and {Naiman}, Jill},
        title = "{First results from the IllustrisTNG simulations: matter and galaxy clustering}",
      journal = {\mnras},
         year = 2018,
        month = mar,
       volume = {475},
       number = {1},
        pages = {676-698},
          doi = {10.1093/mnras/stx3304},
archivePrefix = {arXiv},
       eprint = {1707.03397},
 primaryClass = {astro-ph.GA},
       adsurl = {https://ui.adsabs.harvard.edu/abs/2018MNRAS.475..676S}
}

@ARTICLE{2024MNRAS.532..164L,
       author = {{Lucie-Smith}, Luisa and {Despali}, Giulia and {Springel}, Volker},
        title = "{A deep-learning model for the density profiles of subhaloes in IllustrisTNG}",
      journal = {\mnras},
         year = 2024,
        month = jul,
       volume = {532},
       number = {1},
        pages = {164-176},
          doi = {10.1093/mnras/stae1487},
archivePrefix = {arXiv},
       eprint = {2403.12125},
 primaryClass = {astro-ph.GA},
       adsurl = {https://ui.adsabs.harvard.edu/abs/2024MNRAS.532..164L}
}

@ARTICLE{2023MNRAS.520.5651C,
       author = {{Cannarozzo}, Carlo and {Leauthaud}, Alexie and {Oyarz{\'u}n}, Grecco A. and {Nipoti}, Carlo and {Diemer}, Benedikt and {Huang}, Song and {Rodriguez-Gomez}, Vicente and {Sonnenfeld}, Alessandro and {Bundy}, Kevin},
        title = "{The contribution of in situ and ex situ star formation in early-type galaxies: MaNGA versus IllustrisTNG}",
      journal = {\mnras},
         year = 2023,
        month = apr,
       volume = {520},
       number = {4},
        pages = {5651-5670},
          doi = {10.1093/mnras/stac3023},
archivePrefix = {arXiv},
       eprint = {2210.08109},
 primaryClass = {astro-ph.GA},
       adsurl = {https://ui.adsabs.harvard.edu/abs/2023MNRAS.520.5651C}
}

@ARTICLE{2017AJ....154...86W,
       author = {{Wake}, David A. and {Bundy}, Kevin and {Diamond-Stanic}, Aleksandar M. and {Yan}, Renbin and {Blanton}, Michael R. and {Bershady}, Matthew A. and {S{\'a}nchez-Gallego}, Jos{\'e} R. and {Drory}, Niv and {Jones}, Amy and {Kauffmann}, Guinevere and {Law}, David R. and {Li}, Cheng and {MacDonald}, Nicholas and {Masters}, Karen and {Thomas}, Daniel and {Tinker}, Jeremy and {Weijmans}, Anne-Marie and {Brownstein}, Joel R.},
        title = "{The SDSS-IV MaNGA Sample: Design, Optimization, and Usage Considerations}",
      journal = {\aj},
         year = 2017,
        month = sep,
       volume = {154},
       number = {3},
          eid = {86},
        pages = {86},
          doi = {10.3847/1538-3881/aa7ecc},
archivePrefix = {arXiv},
       eprint = {1707.02989},
 primaryClass = {astro-ph.GA},
       adsurl = {https://ui.adsabs.harvard.edu/abs/2017AJ....154...86W}
}

@ARTICLE{2025A&A...697A.196I,
       author = {{Igo}, Z. and {Merloni}, A.},
        title = "{The global energetics of radio AGN kinetic feedback in the local Universe}",
      journal = {\aap},
         year = 2025,
        month = may,
       volume = {697},
          eid = {A196},
        pages = {A196},
          doi = {10.1051/0004-6361/202452888},
archivePrefix = {arXiv},
       eprint = {2504.00090},
 primaryClass = {astro-ph.GA},
       adsurl = {https://ui.adsabs.harvard.edu/abs/2025A&A...697A.196I}
}

@article{Shreeram_2025,
   title={Retrieving the hot circumgalactic medium physics from the X-ray radial profile from eROSITA with an IlustrisTNG-based forward model},
   volume={703},
   ISSN={1432-0746},
   url={http://dx.doi.org/10.1051/0004-6361/202554508},
   DOI={10.1051/0004-6361/202554508},
   journal={Astronomy \& Astrophysics},
   publisher={EDP Sciences},
   author={Shreeram, Soumya and Comparat, Johan and Merloni, Andrea and Ponti, Gabriele and Popesso, Paola and Zhang, Yi and Nandra, Kirpal and Salvato, Mara and Marini, Ilaria and Buchner, Johannes and Locatelli, Nicola and Igo, Zsofi},
   year={2025},
   month=Nov, pages={A137} }

@ARTICLE{2011ApJ...739L..40R,
       author = {{Rodighiero}, G. and {Daddi}, E. and {Baronchelli}, I. and {Cimatti}, A. and {Renzini}, A. and {Aussel}, H. and {Popesso}, P. and {Lutz}, D. and {Andreani}, P. and {Berta}, S. and {Cava}, A. and {Elbaz}, D. and {Feltre}, A. and {Fontana}, A. and {F{\"o}rster Schreiber}, N.~M. and {Franceschini}, A. and {Genzel}, R. and {Grazian}, A. and {Gruppioni}, C. and {Ilbert}, O. and {Le Floch}, E. and {Magdis}, G. and {Magliocchetti}, M. and {Magnelli}, B. and {Maiolino}, R. and {McCracken}, H. and {Nordon}, R. and {Poglitsch}, A. and {Santini}, P. and {Pozzi}, F. and {Riguccini}, L. and {Tacconi}, L.~J. and {Wuyts}, S. and {Zamorani}, G.},
        title = "{The Lesser Role of Starbursts in Star Formation at z = 2}",
      journal = {\apjl},
         year = 2011,
        month = oct,
       volume = {739},
       number = {2},
          eid = {L40},
        pages = {L40},
          doi = {10.1088/2041-8205/739/2/L40},
archivePrefix = {arXiv},
       eprint = {1108.0933},
 primaryClass = {astro-ph.CO},
       adsurl = {https://ui.adsabs.harvard.edu/abs/2011ApJ...739L..40R}
}

@ARTICLE{2025A&A...693A.197Z,
       author = {{Zhang}, Yi and {Comparat}, Johan and {Ponti}, Gabriele and {Merloni}, Andrea and {Nandra}, Kirpal and {Haberl}, Frank and {Truong}, Nhut and {Pillepich}, Annalisa and {Popesso}, Paola and {Locatelli}, Nicola and {Zhang}, Xiaoyuan and {Sanders}, Jeremy and {Zheng}, Xueying and {Liu}, Ang and {Liu}, Teng and {Predehl}, Peter and {Salvato}, Mara and {Bruggen}, Marcus and {Shreeram}, Soumya and {Yeung}, Michael C.~H.},
        title = "{The hot circumgalactic medium in the eROSITA All-Sky Survey: III. Star-forming and quiescent galaxies}",
      journal = {\aap},
         year = 2025,
        month = jan,
       volume = {693},
          eid = {A197},
        pages = {A197},
          doi = {10.1051/0004-6361/202452273},
archivePrefix = {arXiv},
       eprint = {2411.19945},
 primaryClass = {astro-ph.GA},
       adsurl = {https://ui.adsabs.harvard.edu/abs/2025A&A...693A.197Z}
}

@article{10.1093/mnras/stag1314,
    author = {Bigwood, Leah and Yamamoto, Masaya and Siegel, Jared and Amon, Alexandra and McCarthy, Ian G and Dave, Romeel and Salcido, Jaime and Schaller, Matthieu and Schaye, Joop and Yang, Tianyi},
    title = {The kinetic Sunyaev Zeldovich effect as a benchmark for AGN feedback models in hydrodynamical simulations: insights from DESI + ACT},
    journal = {Monthly Notices of the Royal Astronomical Society},
    pages = {stag1314},
    year = {2026},
    month = {07},
    issn = {0035-8711},
    doi = {10.1093/mnras/stag1314},
    url = {https://doi.org/10.1093/mnras/stag1314},
    eprint = {https://academic.oup.com/mnras/advance-article-pdf/doi/10.1093/mnras/stag1314/68760488/stag1314.pdf},
}

@ARTICLE{2026ApJ..1003..151S,
       author = {{Siegel}, Jared C. and {Amon}, Alexandra and {McCarthy}, Ian G. and {Bigwood}, Leah and {Yamamoto}, Masaya and {Bulbul}, Esra and {Greene}, Jenny E. and {McCullough}, Jamie and {Schaller}, Matthieu and {Schaye}, Joop},
        title = "{Joint X-Ray, Kinetic Sunyaev─Zeldovich, and Weak Lensing Measurements: Toward a Consensus Picture of Efficient Gas Expulsion from Groups and Clusters}",
      journal = {\apj},
         year = 2026,
        month = jun,
       volume = {1003},
       number = {2},
          eid = {151},
        pages = {151},
          doi = {10.3847/1538-4357/ae5dc2},
archivePrefix = {arXiv},
       eprint = {2509.10455},
 primaryClass = {astro-ph.CO},
       adsurl = {https://ui.adsabs.harvard.edu/abs/2026ApJ..1003..151S}
}

@ARTICLE{2025PhRvD.112l3507H,
       author = {{Hadzhiyska}, Boryana and {Ferraro}, Simone and {Farren}, Gerrit S. and {Sailer}, Noah and {Zhou}, Rongpu},
        title = "{Missing baryons recovered: A measurement of the gas fraction in galaxies and groups with the kinematic Sunyaev-Zel'dovich effect and CMB lensing}",
      journal = {\prd},
         year = 2025,
        month = dec,
       volume = {112},
       number = {12},
          eid = {123507},
        pages = {123507},
          doi = {10.1103/mdhz-fgj8},
archivePrefix = {arXiv},
       eprint = {2507.14136},
 primaryClass = {astro-ph.CO},
       adsurl = {https://ui.adsabs.harvard.edu/abs/2025PhRvD.112l3507H}
}

@ARTICLE{Vogelsberger14,
       author = {{Vogelsberger}, Mark and {Genel}, Shy and {Springel}, Volker and {Torrey}, Paul and {Sijacki}, Debora and {Xu}, Dandan and {Snyder}, Greg and {Nelson}, Dylan and {Hernquist}, Lars},
        title = "{Introducing the Illustris Project: simulating the coevolution of dark and visible matter in the Universe}",
      journal = {\mnras},
         year = 2014,
        month = oct,
       volume = {444},
       number = {2},
        pages = {1518-1547},
          doi = {10.1093/mnras/stu1536},
archivePrefix = {arXiv},
       eprint = {1405.2921},
 primaryClass = {astro-ph.CO},
       adsurl = {https://ui.adsabs.harvard.edu/abs/2014MNRAS.444.1518V}
}

@ARTICLE{Schaye15,
       author = {{Schaye}, Joop and {Crain}, Robert A. and {Bower}, Richard G. and {Furlong}, Michelle and {Schaller}, Matthieu and {Theuns}, Tom and {Dalla Vecchia}, Claudio and {Frenk}, Carlos S. and {McCarthy}, I.~G. and {Helly}, John C. and {Jenkins}, Adrian and {Rosas-Guevara}, Y.~M. and {White}, Simon D.~M. and {Baes}, Maarten and {Booth}, C.~M. and {Camps}, Peter and {Navarro}, Julio F. and {Qu}, Yan and {Rahmati}, Alireza and {Sawala}, Till and {Thomas}, Peter A. and {Trayford}, James},
        title = "{The EAGLE project: simulating the evolution and assembly of galaxies and their environments}",
      journal = {\mnras},
         year = 2015,
        month = jan,
       volume = {446},
       number = {1},
        pages = {521-554},
          doi = {10.1093/mnras/stu2058},
archivePrefix = {arXiv},
       eprint = {1407.7040},
 primaryClass = {astro-ph.GA},
       adsurl = {https://ui.adsabs.harvard.edu/abs/2015MNRAS.446..521S}
}

@ARTICLE{McCarthy17,
       author = {{McCarthy}, Ian G. and {Schaye}, Joop and {Bird}, Simeon and {Le Brun}, Amandine M.~C.},
        title = {The BAHAMAS project: calibrated hydrodynamical simulations for large-scale structure cosmology},
      journal = {\mnras},
         year = 2017,
        month = mar,
       volume = {465},
       number = {3},
        pages = {2936-2965},
          doi = {10.1093/mnras/stw2792},
archivePrefix = {arXiv},
       eprint = {1603.02702},
 primaryClass = {astro-ph.CO},
       adsurl = {https://ui.adsabs.harvard.edu/abs/2017MNRAS.465.2936M}
}

@ARTICLE{Dolag16,
       author = {{Dolag}, K. and {Komatsu}, E. and {Sunyaev}, R.},
        title = "{SZ effects in the Magneticum Pathfinder simulation: comparison with the Planck, SPT, and ACT results}",
      journal = {\mnras},
         year = 2016,
        month = dec,
       volume = {463},
       number = {2},
        pages = {1797-1811},
          doi = {10.1093/mnras/stw2035},
archivePrefix = {arXiv},
       eprint = {1509.05134},
 primaryClass = {astro-ph.CO},
       adsurl = {https://ui.adsabs.harvard.edu/abs/2016MNRAS.463.1797D}
}

@article{Schaye:2023jqv,
    author = "Schaye, Joop and others",
    title = "{The FLAMINGO project: cosmological hydrodynamical simulations for large-scale structure and galaxy cluster surveys}",
    eprint = "2306.04024",
    archivePrefix = "arXiv",
    primaryClass = "astro-ph.CO",
    doi = "10.1093/mnras/stad2419",
    journal = "Mon. Not. Roy. Astron. Soc.",
    volume = "526",
    number = "4",
    pages = "4978--5020",
    year = "2023"
}

@ARTICLE{dolag25,
       author = {{Dolag}, Klaus and {Remus}, Rhea-Silvia and {Valenzuela}, Lucas M. and {Kimmig}, Lucas C. and {Seidel}, Benjamin and {Fortune}, Silvio and {Stoiber}, Johannes and {Ivleva}, Anna and {Hoffmann}, Tadziu and {Biffi}, Veronica and {Marini}, Ilaria and {Popesso}, Paola and {Vladutescu-Zopp}, Stephan},
        title = "{Encyclopedia Magneticum: Scaling Relations from Cosmic Dawn to Present Day}",
      journal = {\aap},
         year = 2025,
        month = apr,
          eid = {arXiv:2504.01061},
        pages = {submitted},
          doi = {10.48550/arXiv.2504.01061},
archivePrefix = {arXiv},
       eprint = {2504.01061},
 primaryClass = {astro-ph.CO},
       adsurl = {https://ui.adsabs.harvard.edu/abs/2025arXiv250401061D}
}

@article{article,
author = {Lanman, Adam and Simha, Sunil and Masui, Kiyoshi and Prochaska, J. and Darlinger, Rachel and Dong, Fengqiu and Gaensler, Bryan and Joseph, Ronniy and Kaczmarek, Jane and Kahinga, Lordrick and Khan, Afrokk and Leung, Calvin and Mas-Ribas, Lluis and Patil, Swarali and Pearlman, Aaron and Sammons, Mawson and Shin, Kaitlyn and Smith, Kendrick and Wang, Haochen},
year = {2026},
month = {05},
pages = {5},
title = {Constraining Gas Mass Fractions in Galaxy Groups and Clusters with the First CHIME/FRB Outrigger},
volume = {1003},
journal = {The Astrophysical Journal},
doi = {10.3847/1538-4357/ae606e}
}

@ARTICLE{angles17,
       author = {{Angl{\'e}s-Alc{\'a}zar}, Daniel and {Dav{\'e}}, Romeel and {Faucher-Gigu{\`e}re}, Claude-Andr{\'e} and {{\"O}zel}, Feryal and {Hopkins}, Philip F.},
        title = "{Gravitational torque-driven black hole growth and feedback in cosmological simulations}",
      journal = {\mnras},
         year = 2017,
        month = jan,
       volume = {464},
       number = {3},
        pages = {2840-2853},
          doi = {10.1093/mnras/stw2565},
archivePrefix = {arXiv},
       eprint = {1603.08007},
 primaryClass = {astro-ph.GA},
       adsurl = {https://ui.adsabs.harvard.edu/abs/2017MNRAS.464.2840A}
}

@ARTICLE{Best2012,
       author = {{Best}, P.~N. and {Heckman}, T.~M.},
        title = "{On the fundamental dichotomy in the local radio-AGN population: accretion, evolution and host galaxy properties}",
      journal = {\mnras},
         year = 2012,
        month = apr,
       volume = {421},
       number = {2},
        pages = {1569-1582},
          doi = {10.1111/j.1365-2966.2012.20414.x},
archivePrefix = {arXiv},
       eprint = {1201.2397},
 primaryClass = {astro-ph.CO},
       adsurl = {https://ui.adsabs.harvard.edu/abs/2012MNRAS.421.1569B}
}

@ARTICLE{Sanchez2022,
       author = {{S{\'a}nchez}, S.~F. and {Barrera-Ballesteros}, J.~K. and {Lacerda}, E. and {Mej{\'\i}a-Narvaez}, A. and {Camps-Fari{\~n}a}, A. and {Bruzual}, Gustavo and {Espinosa-Ponce}, C. and {Rodr{\'\i}guez-Puebla}, A. and {Calette}, A.~R. and {Ibarra-Medel}, H. and {Avila-Reese}, V. and {Hernandez-Toledo}, H. and {Bershady}, M.~A. and {Cano-Diaz}, M. and {Munguia-Cordova}, A.~M.},
        title = "{SDSS-IV MaNGA: pyPipe3D Analysis Release for 10,000 Galaxies}",
      journal = {\apjs},
         year = 2022,
        month = oct,
       volume = {262},
       number = {2},
          eid = {36},
        pages = {36},
          doi = {10.3847/1538-4365/ac7b8f},
archivePrefix = {arXiv},
       eprint = {2206.07062},
 primaryClass = {astro-ph.GA},
       adsurl = {https://ui.adsabs.harvard.edu/abs/2022ApJS..262...36S}
}

@ARTICLE{Lyskova23,
       author = {{Lyskova}, N. and {Churazov}, E. and {Khabibullin}, I.~I. and {Burenin}, R. and {Starobinsky}, A.~A. and {Sunyaev}, R.},
        title = "{X-ray surface brightness and gas density profiles of galaxy clusters up to 3 {\texttimes} R$_{500c}$ with SRG/eROSITA}",
      journal = {\mnras},
         year = 2023,
        month = oct,
       volume = {525},
       number = {1},
        pages = {898-907},
          doi = {10.1093/mnras/stad2305},
archivePrefix = {arXiv},
       eprint = {2305.07080},
 primaryClass = {astro-ph.CO},
       adsurl = {https://ui.adsabs.harvard.edu/abs/2023MNRAS.525..898L}
}

@ARTICLE{2023A&A...673A..16K,
       author = {{Kouroumpatzakis}, K. and {Zezas}, A. and {Kyritsis}, E. and {Salim}, S. and {Svoboda}, J.},
        title = "{Star formation rate and stellar mass calibrations based on infrared photometry and their dependence on stellar population age and extinction}",
      journal = {\aap},
         year = 2023,
        month = may,
       volume = {673},
          eid = {A16},
        pages = {A16},
          doi = {10.1051/0004-6361/202346054},
archivePrefix = {arXiv},
       eprint = {2303.10013},
 primaryClass = {astro-ph.GA},
       adsurl = {https://ui.adsabs.harvard.edu/abs/2023A&A...673A..16K}
}

@ARTICLE{2007ApJS..173..267S,
       author = {{Salim}, Samir and {Rich}, R. Michael and {Charlot}, St{\'e}phane and {Brinchmann}, Jarle and {Johnson}, Benjamin D. and {Schiminovich}, David and {Seibert}, Mark and {Mallery}, Ryan and {Heckman}, Timothy M. and {Forster}, Karl and {Friedman}, Peter G. and {Martin}, D. Christopher and {Morrissey}, Patrick and {Neff}, Susan G. and {Small}, Todd and {Wyder}, Ted K. and {Bianchi}, Luciana and {Donas}, Jos{\'e} and {Lee}, Young-Wook and {Madore}, Barry F. and {Milliard}, Bruno and {Szalay}, Alex S. and {Welsh}, Barry Y. and {Yi}, Sukyoung K.},
        title = "{UV Star Formation Rates in the Local Universe}",
      journal = {\apjs},
         year = 2007,
        month = dec,
       volume = {173},
       number = {2},
        pages = {267-292},
          doi = {10.1086/519218},
archivePrefix = {arXiv},
       eprint = {0704.3611},
 primaryClass = {astro-ph},
       adsurl = {https://ui.adsabs.harvard.edu/abs/2007ApJS..173..267S}
}

@ARTICLE{2015ApJ...799..125V,
       author = {{van de Sande}, Jesse and {Kriek}, Mariska and {Franx}, Marijn and {Bezanson}, Rachel and {van Dokkum}, Pieter G.},
        title = "{The Relation between Dynamical Mass-to-light Ratio and Color for Massive Quiescent Galaxies out to z \raisebox{-0.5ex}\textasciitilde 2 and Comparison with Stellar Population Synthesis Models}",
      journal = {\apj},
         year = 2015,
        month = feb,
       volume = {799},
       number = {2},
          eid = {125},
        pages = {125},
          doi = {10.1088/0004-637X/799/2/125},
archivePrefix = {arXiv},
       eprint = {1411.5363},
 primaryClass = {astro-ph.GA},
       adsurl = {https://ui.adsabs.harvard.edu/abs/2015ApJ...799..125V}
}

@ARTICLE{2022ApJ...927..164B,
       author = {{Borghi}, Nicola and {Moresco}, Michele and {Cimatti}, Andrea and {Huchet}, Alexandre and {Quai}, Salvatore and {Pozzetti}, Lucia},
        title = "{Toward a Better Understanding of Cosmic Chronometers: Stellar Population Properties of Passive Galaxies at Intermediate Redshift}",
      journal = {\apj},
         year = 2022,
        month = mar,
       volume = {927},
       number = {2},
          eid = {164},
        pages = {164},
          doi = {10.3847/1538-4357/ac3240},
archivePrefix = {arXiv},
       eprint = {2106.14894},
 primaryClass = {astro-ph.GA},
       adsurl = {https://ui.adsabs.harvard.edu/abs/2022ApJ...927..164B}
}

@ARTICLE{2013MNRAS.428.3121M,
       author = {{Moster}, Benjamin P. and {Naab}, Thorsten and {White}, Simon D.~M.},
        title = "{Galactic star formation and accretion histories from matching galaxies to dark matter haloes}",
      journal = {\mnras},
         year = 2013,
        month = feb,
       volume = {428},
       number = {4},
        pages = {3121-3138},
          doi = {10.1093/mnras/sts261},
archivePrefix = {arXiv},
       eprint = {1205.5807},
 primaryClass = {astro-ph.CO},
       adsurl = {https://ui.adsabs.harvard.edu/abs/2013MNRAS.428.3121M}
}

@ARTICLE{2018MNRAS.477.1822M,
       author = {{Moster}, Benjamin P. and {Naab}, Thorsten and {White}, Simon D.~M.},
        title = "{EMERGE - an empirical model for the formation of galaxies since z {\ensuremath{\sim}} 10}",
      journal = {\mnras},
         year = 2018,
        month = jun,
       volume = {477},
       number = {2},
        pages = {1822-1852},
          doi = {10.1093/mnras/sty655},
archivePrefix = {arXiv},
       eprint = {1705.05373},
 primaryClass = {astro-ph.GA},
       adsurl = {https://ui.adsabs.harvard.edu/abs/2018MNRAS.477.1822M}
}

@ARTICLE{2017MNRAS.470..651R,
       author = {{Rodr{\'\i}guez-Puebla}, Aldo and {Primack}, Joel R. and {Avila-Reese}, Vladimir and {Faber}, S.~M.},
        title = "{Constraining the galaxy-halo connection over the last 13.3 Gyr: star formation histories, galaxy mergers and structural properties}",
      journal = {\mnras},
         year = 2017,
        month = sep,
       volume = {470},
       number = {1},
        pages = {651-687},
          doi = {10.1093/mnras/stx1172},
archivePrefix = {arXiv},
       eprint = {1703.04542},
 primaryClass = {astro-ph.GA},
       adsurl = {https://ui.adsabs.harvard.edu/abs/2017MNRAS.470..651R}
}

@ARTICLE{2014ApJ...793...12B,
       author = {{Birrer}, Simon and {Lilly}, Simon and {Amara}, Adam and {Paranjape}, Aseem and {Refregier}, Alexandre},
        title = "{A Simple Model Linking Galaxy and Dark Matter Evolution}",
      journal = {\apj},
         year = 2014,
        month = sep,
       volume = {793},
       number = {1},
          eid = {12},
        pages = {12},
          doi = {10.1088/0004-637X/793/1/12},
archivePrefix = {arXiv},
       eprint = {1401.3162},
 primaryClass = {astro-ph.CO},
       adsurl = {https://ui.adsabs.harvard.edu/abs/2014ApJ...793...12B}
}

@ARTICLE{2018AstL...44....8K,
       author = {{Kravtsov}, A.~V. and {Vikhlinin}, A.~A. and {Meshcheryakov}, A.~V.},
        title = "{Stellar Mass{\textemdash}Halo Mass Relation and Star Formation Efficiency in High-Mass Halos}",
      journal = {Astronomy Letters},
         year = 2018,
        month = jan,
       volume = {44},
       number = {1},
        pages = {8-34},
          doi = {10.1134/S1063773717120015},
archivePrefix = {arXiv},
       eprint = {1401.7329},
 primaryClass = {astro-ph.CO},
       adsurl = {https://ui.adsabs.harvard.edu/abs/2018AstL...44....8K}
}

@ARTICLE{2019A&A...631A.175E,
       author = {{Erfanianfar}, G. and {Finoguenov}, A. and {Furnell}, K. and {Popesso}, P. and {Biviano}, A. and {Wuyts}, S. and {Collins}, C.~A. and {Mirkazemi}, M. and {Comparat}, J. and {Khosroshahi}, H. and {Nandra}, K. and {Capasso}, R. and {Rykoff}, E. and {Wilman}, D. and {Merloni}, A. and {Clerc}, N. and {Salvato}, M. and {Chitham}, J.~I. and {Kelvin}, L.~S. and {Gozaliasl}, G. and {Weijmans}, A. and {Brownstein}, J. and {Egami}, E. and {Pereira}, M.~J. and {Schneider}, D.~P. and {Kirkpatrick}, C. and {Damsted}, S. and {Kukkola}, A.},
        title = "{Stellar mass-halo mass relation for the brightest central galaxies of X-ray clusters since z {\ensuremath{\sim}} 0.65}",
      journal = {\aap},
         year = 2019,
        month = nov,
       volume = {631},
          eid = {A175},
        pages = {A175},
          doi = {10.1051/0004-6361/201935375},
archivePrefix = {arXiv},
       eprint = {1908.01559},
 primaryClass = {astro-ph.GA},
       adsurl = {https://ui.adsabs.harvard.edu/abs/2019A&A...631A.175E}
}

@ARTICLE{2022ApJ...928...28G,
       author = {{Golden-Marx}, Jesse B. and {Miller}, C.~J. and {Zhang}, Y. and {Ogando}, R.~L.~C. and {Palmese}, A. and {Abbott}, T.~M.~C. and {Aguena}, M. and {Allam}, S. and {Andrade-Oliveira}, F. and {Annis}, J. and {Bacon}, D. and {Bertin}, E. and {Brooks}, D. and {Buckley-Geer}, E. and {Carnero Rosell}, A. and {Carrasco Kind}, M. and {Castander}, F.~J. and {Costanzi}, M. and {Crocce}, M. and {da Costa}, L.~N. and {Pereira}, M.~E.~S. and {De Vicente}, J. and {Desai}, S. and {Diehl}, H.~T. and {Doel}, P. and {Drlica-Wagner}, A. and {Everett}, S. and {Evrard}, A.~E. and {Ferrero}, I. and {Flaugher}, B. and {Fosalba}, P. and {Frieman}, J. and {Garc{\'\i}a-Bellido}, J. and {Gaztanaga}, E. and {Gerdes}, D.~W. and {Gruen}, D. and {Gruendl}, R.~A. and {Gschwend}, J. and {Gutierrez}, G. and {Hartley}, W.~G. and {Hinton}, S.~R. and {Hollowood}, D.~L. and {Honscheid}, K. and {Hoyle}, B. and {James}, D.~J. and {Jeltema}, T. and {Kim}, A.~G. and {Krause}, E. and {Kuehn}, K. and {Kuropatkin}, N. and {Lahav}, O. and {Lima}, M. and {Maia}, M.~A.~G. and {Marshall}, J.~L. and {Melchior}, P. and {Menanteau}, F. and {Miquel}, R. and {Mohr}, J.~J. and {Morgan}, R. and {Paz-Chinch{\'o}n}, F. and {Petravick}, D. and {Pieres}, A. and {Plazas Malag{\'o}n}, A.~A. and {Prat}, J. and {Romer}, A.~K. and {Sanchez}, E. and {Santiago}, B. and {Scarpine}, V. and {Schubnell}, M. and {Serrano}, S. and {Sevilla-Noarbe}, I. and {Smith}, M. and {Soares-Santos}, M. and {Suchyta}, E. and {Tarle}, G. and {Varga}, T.~N.},
        title = "{The Observed Evolution of the Stellar Mass-Halo Mass Relation for Brightest Central Galaxies}",
      journal = {\apj},
         year = 2022,
        month = mar,
       volume = {928},
       number = {1},
          eid = {28},
        pages = {28},
          doi = {10.3847/1538-4357/ac4cb4},
archivePrefix = {arXiv},
       eprint = {2107.02197},
 primaryClass = {astro-ph.GA},
       adsurl = {https://ui.adsabs.harvard.edu/abs/2022ApJ...928...28G}
}

@ARTICLE{2015MNRAS.450.1604L,
       author = {{Lu}, Zhankui and {Mo}, H.~J. and {Lu}, Yu and {Katz}, Neal and {Weinberg}, Martin D. and {van den Bosch}, Frank C. and {Yang}, Xiaohu},
        title = "{Star formation and stellar mass assembly in dark matter haloes: from giants to dwarfs}",
      journal = {\mnras},
         year = 2015,
        month = jun,
       volume = {450},
       number = {2},
        pages = {1604-1617},
          doi = {10.1093/mnras/stv667},
archivePrefix = {arXiv},
       eprint = {1406.5068},
 primaryClass = {astro-ph.GA},
       adsurl = {https://ui.adsabs.harvard.edu/abs/2015MNRAS.450.1604L}
}

@ARTICLE{2003ApJ...586L.133C,
       author = {{Chabrier}, Gilles},
        title = "{The Galactic Disk Mass Function: Reconciliation of the Hubble Space Telescope and Nearby Determinations}",
      journal = {\apjl},
         year = 2003,
        month = apr,
       volume = {586},
       number = {2},
        pages = {L133-L136},
          doi = {10.1086/374879},
archivePrefix = {arXiv},
       eprint = {astro-ph/0302511},
 primaryClass = {astro-ph},
       adsurl = {https://ui.adsabs.harvard.edu/abs/2003ApJ...586L.133C}
}

@ARTICLE{2017MNRAS.465.3291W,
       author = {{Weinberger}, Rainer and {Springel}, Volker and {Hernquist}, Lars and {Pillepich}, Annalisa and {Marinacci}, Federico and {Pakmor}, R{\"u}diger and {Nelson}, Dylan and {Genel}, Shy and {Vogelsberger}, Mark and {Naiman}, Jill and {Torrey}, Paul},
        title = "{Simulating galaxy formation with black hole driven thermal and kinetic feedback}",
      journal = {\mnras},
         year = 2017,
        month = mar,
       volume = {465},
       number = {3},
        pages = {3291-3308},
          doi = {10.1093/mnras/stw2944},
archivePrefix = {arXiv},
       eprint = {1607.03486},
 primaryClass = {astro-ph.GA},
       adsurl = {https://ui.adsabs.harvard.edu/abs/2017MNRAS.465.3291W}
}

@ARTICLE{2013MNRAS.428.2966P,
       author = {{Puchwein}, Ewald and {Springel}, Volker},
        title = "{Shaping the galaxy stellar mass function with supernova- and AGN-driven winds}",
      journal = {\mnras},
         year = 2013,
        month = feb,
       volume = {428},
       number = {4},
        pages = {2966-2979},
          doi = {10.1093/mnras/sts243},
archivePrefix = {arXiv},
       eprint = {1205.2694},
 primaryClass = {astro-ph.CO},
       adsurl = {https://ui.adsabs.harvard.edu/abs/2013MNRAS.428.2966P}
}

@ARTICLE{2022ApJ...937..117F,
       author = {{Fraser-McKelvie}, A. and {Cortese}, L.},
        title = "{Beyond Galaxy Bimodality: The Complex Interplay between Kinematic Morphology and Star Formation in the Local Universe}",
      journal = {\apj},
         year = 2022,
        month = oct,
       volume = {937},
       number = {2},
          eid = {117},
        pages = {117},
          doi = {10.3847/1538-4357/ac874d},
archivePrefix = {arXiv},
       eprint = {2208.01936},
 primaryClass = {astro-ph.GA},
       adsurl = {https://ui.adsabs.harvard.edu/abs/2022ApJ...937..117F}
}

@article{Mulcahey_2022,
	author = {{Mulcahey, C. R.} and {Leslie, S. K.} and {Jackson, T. M.} and {Young, J. E.} and {Prandoni, I.} and {Hardcastle, M. J.} and {Roy, N.} and {Małek, K.} and {Magliocchetti, M.} and {Bonato, M.} and {Röttgering, H. J. A.} and {Drabent, A.}},
	title = {Star formation and AGN feedback in the local Universe: Combining LOFAR and MaNGA⋆},
	DOI= "10.1051/0004-6361/202142215",
	url= "https://doi.org/10.1051/0004-6361/202142215",
	journal = {Astronomy and Astrophysics},
	year = 2022,
	volume = 665,
	pages = "A144",
}

@article{best2012fundamental,
  title={On the fundamental dichotomy in the local radio-AGN population: accretion, evolution and host galaxy properties},
  author={Best, PN and Heckman, TM},
  journal={Monthly Notices of the Royal Astronomical Society},
  volume={421},
  number={2},
  pages={1569--1582},
  year={2012},
  publisher={Blackwell Publishing Ltd Oxford, UK}
}

@article{gurkan2018lofar,
  title={LOFAR/H-ATLAS: the low-frequency radio luminosity--star formation rate relation},
  author={G{\"u}rkan, G{\"u}lay and Hardcastle, Martin J and Smith, Dan JB and Best, Philip N and Bourne, Nathan and Calistro-Rivera, Gabriela and Heald, George and Jarvis, Matt J and Prandoni, Isabella and R{\"o}ttgering, HJA and others},
  journal={Monthly Notices of the Royal Astronomical Society},
  volume={475},
  number={3},
  pages={3010--3028},
  year={2018},
  publisher={Oxford University Press}
}

@article{smith2021lofar,
  title={The LOFAR Two-metre Sky Survey Deep Fields-The star-formation rate--radio luminosity relation at low frequencies},
  author={Smith, DJB and Haskell, P and G{\"u}rkan, G and Best, PN and Hardcastle, MJ and Kondapally, R and Williams, W and Duncan, KJ and Cochrane, RK and McCheyne, I and others},
  journal={Astronomy \& Astrophysics},
  volume={648},
  pages={A6},
  year={2021},
  publisher={EDP Sciences}
}

@article{best2023lofar,
  title={The LOFAR Two-metre Sky Survey: Deep Fields data release 1. V. Survey description, source classifications, and host galaxy properties},
  author={Best, PN and Kondapally, Rohit and Williams, WL and Cochrane, RK and Duncan, KJ and Hale, CL and Haskell, P and Ma{\l}ek, K and McCheyne, I and Smith, DJB and others},
  journal={Monthly Notices of the Royal Astronomical Society},
  volume={523},
  number={2},
  pages={1729--1755},
  year={2023},
  publisher={Oxford University Press}
}

@article{hardcastle2023lofar,
  title={The LOFAR Two-Metre Sky Survey-VI. Optical identifications for the second data release},
  author={Hardcastle, MJ and Horton, MA and Williams, WL and Duncan, KJ and Alegre, L and Barkus, B and Croston, JH and Dickinson, H and Osinga, E and R{\"o}ttgering, HJA and others},
  journal={Astronomy \& Astrophysics},
  volume={678},
  pages={A151},
  year={2023},
  publisher={EDP Sciences}
}

@misc{chilufya2024naturecompactradioloudagn,
      title={The nature of compact radio-loud AGN: a systematic look at the LOFAR AGN population}, 
      author={J. Chilufya and M. J. Hardcastle and J. C. S. Pierce and J. H. Croston and B. Mingo and X. Zheng and R. D. Baldi and H. J. A. Röttgering},
      year={2024},
      eprint={2402.19424},
      archivePrefix={arXiv},
      primaryClass={astro-ph.GA},
      url={https://arxiv.org/abs/2402.19424}, 
}

@ARTICLE{1992ARA&A..30..575C,
       author = {{Condon}, J.~J.},
        title = "{Radio emission from normal galaxies.}",
      journal = {\araa},
         year = 1992,
        month = jan,
       volume = {30},
        pages = {575-611},
          doi = {10.1146/annurev.aa.30.090192.003043},
       adsurl = {https://ui.adsabs.harvard.edu/abs/1992ARA&A..30..575C}
}

@ARTICLE{2021MNRAS.500.4004D,
       author = {{Donnari}, Martina and {Pillepich}, Annalisa and {Joshi}, Gandhali D. and {Nelson}, Dylan and {Genel}, Shy and {Marinacci}, Federico and {Rodriguez-Gomez}, Vicente and {Pakmor}, R{\"u}diger and {Torrey}, Paul and {Vogelsberger}, Mark and {Hernquist}, Lars},
        title = "{Quenched fractions in the IllustrisTNG simulations: the roles of AGN feedback, environment, and pre-processing}",
      journal = {\mnras},
         year = 2021,
        month = jan,
       volume = {500},
       number = {3},
        pages = {4004-4024},
          doi = {10.1093/mnras/staa3006},
archivePrefix = {arXiv},
       eprint = {2008.00005},
 primaryClass = {astro-ph.GA},
       adsurl = {https://ui.adsabs.harvard.edu/abs/2021MNRAS.500.4004D}
}

@ARTICLE{2016ApJS..227....2S,
       author = {{Salim}, Samir and {Lee}, Janice C. and {Janowiecki}, Steven and {da Cunha}, Elisabete and {Dickinson}, Mark and {Boquien}, M{\'e}d{\'e}ric and {Burgarella}, Denis and {Salzer}, John J. and {Charlot}, St{\'e}phane},
        title = "{GALEX-SDSS-WISE Legacy Catalog (GSWLC): Star Formation Rates, Stellar Masses, and Dust Attenuations of 700,000 Low-redshift Galaxies}",
      journal = {\apjs},
         year = 2016,
        month = nov,
       volume = {227},
       number = {1},
          eid = {2},
        pages = {2},
          doi = {10.3847/0067-0049/227/1/2},
archivePrefix = {arXiv},
       eprint = {1610.00712},
 primaryClass = {astro-ph.GA},
       adsurl = {https://ui.adsabs.harvard.edu/abs/2016ApJS..227....2S}
}

@ARTICLE{2024A&A...686A..43I,
       author = {{Igo}, Z. and {Merloni}, A. and {Hoang}, D. and {Buchner}, J. and {Liu}, T. and {Salvato}, M. and {Arcodia}, R. and {Bellstedt}, S. and {Br{\"u}ggen}, M. and {Croston}, J.~H. and {de Gasperin}, F. and {Georgakakis}, A. and {Hardcastle}, M.~J. and {Nandra}, K. and {Ni}, Q. and {Pasini}, T. and {Shimwell}, T. and {Wolf}, J.},
        title = "{The LOFAR - eFEDS survey: The incidence of radio and X-ray AGN and the disk-jet connection}",
      journal = {\aap},
         year = 2024,
        month = jun,
       volume = {686},
          eid = {A43},
        pages = {A43},
          doi = {10.1051/0004-6361/202349069},
archivePrefix = {arXiv},
       eprint = {2402.16943},
 primaryClass = {astro-ph.HE},
       adsurl = {https://ui.adsabs.harvard.edu/abs/2024A&A...686A..43I}
}

@article{Lammers_2023,
   title={Active Galactic Nuclei Feedback in SDSS-IV MaNGA: AGNs Have Suppressed Central Star Formation Rates},
   volume={953},
   ISSN={1538-4357},
   url={http://dx.doi.org/10.3847/1538-4357/acdd57},
   DOI={10.3847/1538-4357/acdd57},
   number={1},
   journal={The Astrophysical Journal},
   publisher={American Astronomical Society},
   author={Lammers, Caleb and Iyer, Kartheik G. and Ibarra-Medel, Hector and Pacifici, Camilla and Sánchez, Sebastián F. and Tacchella, Sandro and Woo, Joanna},
   year={2023},
   month=jul, pages={26} }

\begin{appendix}
\section{Supplementary information and figures}
This appendix describes different supplementary figures, table, and information.

\subsection{The $f_{\mathrm{gas}}$--$M_{\mathrm{h}}$ relation}
As discussed in the main text, current cosmological simulations show varying levels of agreement with the observed hot gas content of haloes. In this study, we assess whether these same models can also reproduce the observed properties of galaxies.
\begin{figure}[h!]
\centering
\includegraphics[width=1\hsize]{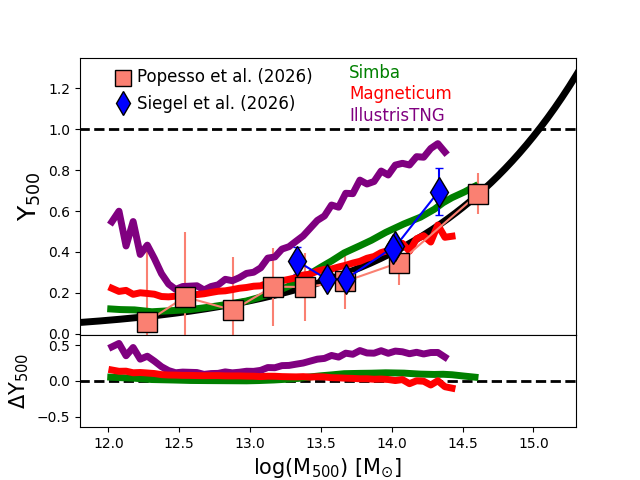}
\caption{Observed and simulated hot gas fractions ($Y_\mathrm{gas}$) normalized to cosmic baryon fraction versus $M_{500}$. Data include Magneticum, SIMBA, and IllustrisTNG \citep{Daves_24}, compared against eROSITA stacking \citep{refId0}, \citep{2026ApJ..1003..151S} and CHEX-MATE clusters \citep{2025A&A...693A...2S, Lyskova23}. The solid curve represents the best fit from \citet{refId0}. Bottom panels show residuals relative to this fit.}
\label{fig:fgas}
\end{figure}
\subsection{Selection of low-luminosity Radio AGN}
\label{app:radioAGN}
\begin{figure}[ht]
\centering
\includegraphics[width=0.45\textwidth]{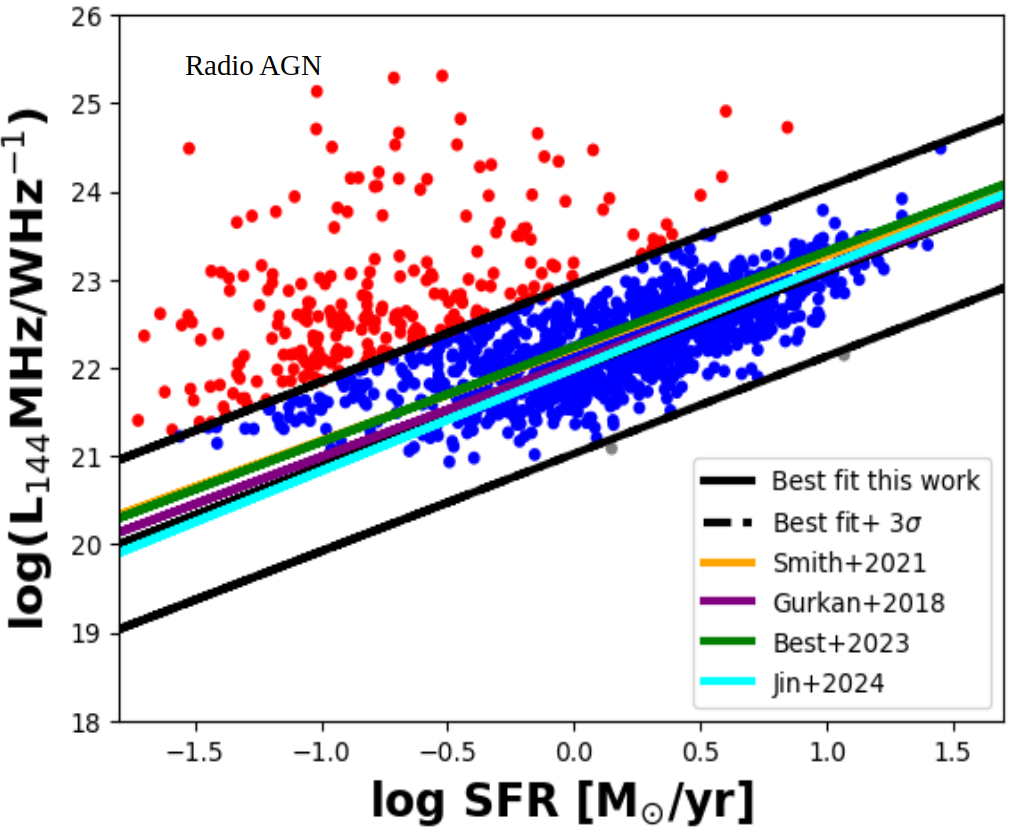}
\caption{Selection of radio AGN based on radio excess relative to the SFR--$L_{144}$ relation. The top and bottom black lines represent $\pm 3\sigma$ of our best fit. Blue points indicate star-forming galaxies, red points denote the 261 radio AGN detected at $+3\sigma$ excess criterion \citep[e.g.,][]{best2012fundamental, gurkan2018lofar, smith2021lofar, best2023lofar} of equation~\ref{eq:sfr_l144}. The resulting radio luminosities span $21.0 \lesssim \log(L_{144}/\mathrm{W\,Hz^{-1}}) \lesssim 25.5$, with a mean value of $\langle \log L_{144} \rangle \simeq 22.8\mathrm{W\,Hz^{-1}}$.}
\label{fig:radioAGN}
\end{figure}

\begin{table*}[t]
\centering
\caption{Summary of the selected AGN catalogues.}
\label{tab:agn_summary}

\begin{tabular}{lccc}
\hline\hline
Catalogue & Initial Size & Final Size & Notes \\
\hline
$S_{\rm AGN}$ (Optical) & 227 & 77 &
Pipe3D-based BPT \citep{Sanchez2022} \\

$C_{\rm AGN}$ & 406 & 172 &
Multiwavelength \citep{Comerford_2020,2024ApJ...963...53C} \\

LoFAR (LoTSS) & 261 & 87 &
Selected via $3\sigma$ radio excess at 144~MHz \\

\hline
Combined AGN & 894 & 336 &
No duplicates \\
\hline
\end{tabular}

\vspace{1mm}
\raggedright
Notes. The sample is selected with
$M_{\star}>10^{10}\,M_{\odot}$ and $z<0.085$ for
Fig.~\ref{fig:plot1}. $C_{\rm AGN}$ comprises X-ray (10), MIR (17),
broad-emission-line (26), and radio AGN [HERG (65), LERG (54)].

\end{table*}
Radio luminosities at 144~MHz were computed using the following relation:
\begin{equation}
\log\left( L_{144}\,[\mathrm{W\,Hz^{-1}}] \right) =
\log \left[ 4\pi D_{\mathrm{L}}^{2}(z)\,F_{144}\,(1+z)^{\alpha-1} \right],
\end{equation}
where $D_{\mathrm{L}}$ is the luminosity distance, and $F_{144}$ is the integrated 144~MHz flux density from LoTSS, and we adopt a fixed spectral index of $\alpha = 0.7$ \citep{1992ARA&A..30..575C}. Based on the cross-matched sample of MaNGA+LoTSS, we derived a SFR--$L_{144}$ relation defined by: 
\begin{equation}
\log(\mathrm{SFR}/M_{\odot}s^{-1}) = (1.105 \pm 0.026)\,\log(L_{144}) + (21.985 \pm 0.044),
\label{eq:sfr_l144}
\end{equation}
with an intrinsic scatter of $\sigma = 0.32$~dex, consistent with previous LoFAR-based studies \citep[e.g.,][]{smith2021lofar}. Figure~\ref{fig:radioAGN} illustrates the radio AGN selection in the SFR--$L_{144}$ plane.

\subsection{Fraction of galaxy populations in the SFR--$M_{\star}$ plane}

Table~\ref{table:stats} summarises the distribution of galaxy populations relative to the Po19 main sequence, complementing the $\Delta\log(\mathrm{SFR})$ distributions, Fig.~\ref{fig:MS}. Despite variations in sample size ($N\approx10^{2}$--$10^{5}$), the qualitative trends are broadly consistent across simulations. In the highest-mass bin ($\log(M_{\star}/M_{\odot})>11$), the number of galaxies ranges from 166 in Box4 to 48,721 in Box2, with MaNGA, SIMBA, and TNG100 containing 1038, 735, and 654 galaxies, respectively. While the smaller samples are more susceptible to statistical uncertainties and cosmic variance, the agreement between the Box2 and Box4 population fractions, despite their substantially different volumes suggest that the main conclusions in this study are robust.

\begin{table*}[t]
\caption{Distribution of MaNGA and simulated galaxies in the SFR-$M_{\star}$ plane per $M_{\star}$ bin. Supplementing Fig.~\ref{fig:MS}}
\label{table:stats}
\centering
\begin{tabular}{l l r r r r r}
\toprule
log ($M_{\star}/\,M_{\odot}$) bin & Dataset & N Total & MS \% & GV \% & RS \% & SB \% \\
\midrule

\multirow{5}{*}{10.0--10.5} 
& MaNGA  & 2287   & 41.23 & 29.25 & 29.47 & 0.04 \\
& Box2   & 400080 & 46.52 & 7.62  & 43.52 & 2.33 \\
& Box4   & 1035   & 17.68 & 12.17 & 69.66 & 0.48 \\
& SIMBA  & 6840   & 23.20 & 21.13 & 54.37 & 1.30 \\
& TNG100 & 3612   & 61.88 & 13.15 & 23.62 & 1.36 \\
\midrule

\multirow{5}{*}{10.5--11.0} 
& MaNGA  & 2560   & 30.74 & 29.34 & 39.61 & 0.31 \\
& Box2   & 205987 & 3.01  & 0.61  & 95.02 & 1.35 \\
& Box4   & 466    & 9.66  & 6.44  & 83.26 & 0.64 \\
& SIMBA  & 2583   & 22.76 & 14.21 & 59.97 & 3.06 \\
& TNG100 & 2171   & 27.73 & 13.77 & 55.73 & 2.76 \\
\midrule

\multirow{5}{*}{>11.0} 
& MaNGA  & 1038   & 15.80 & 28.61 & 55.11 & 0.48 \\
& Box2   & 48721  & 3.46  & 2.79  & 93.46 & 0.28 \\
& Box4   & 166    & 22.89 & 12.65 & 61.45 & 3.01 \\
& SIMBA  & 735    & 15.78 & 6.26  & 76.46 & 1.50 \\
& TNG100 & 654    & 27.22 & 27.22 & 44.50 & 1.07 \\
\bottomrule
\end{tabular}
\end{table*}

\subsubsection{Robustness to the adopted galaxy-classification scheme}\label{robust}
\begin{figure*}[htbp!]
  \centering
\includegraphics[width=0.96\textwidth]
{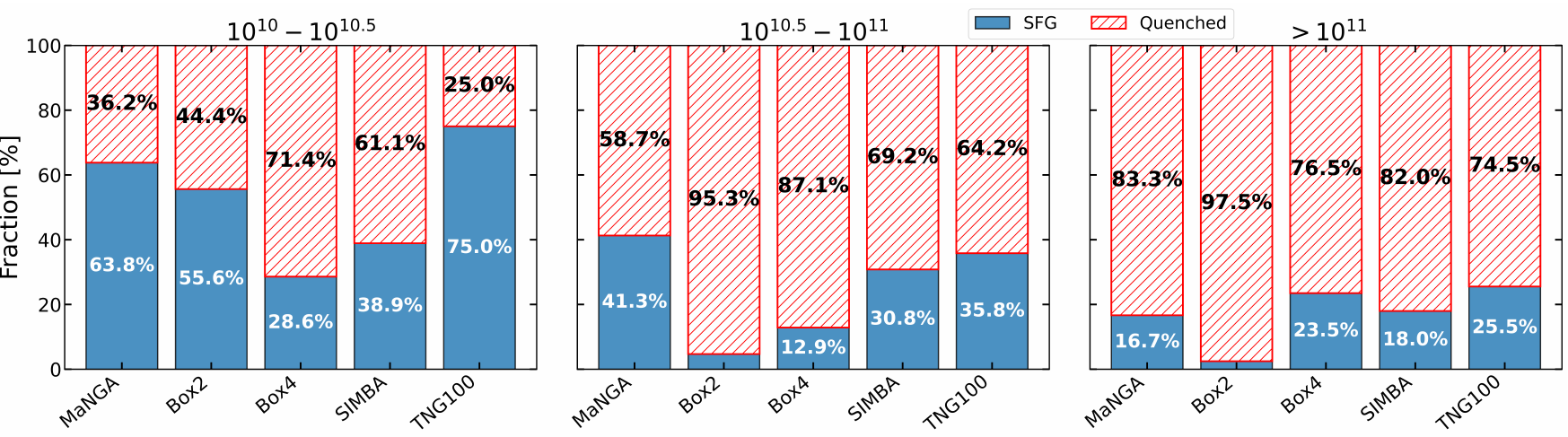}
\caption{Fractions of star-forming galaxies (blue filled bars) and quenched galaxies (red hatched bars) in three stellar-mass bins, $10^{10} \leq M_\star/M_\odot < 10^{10.5}$, $10^{10.5} \leq M_\star/M_\odot < 10^{11}$, and $M_\star/M_\odot > 10^{11}$. Galaxies are classified as star forming for $\log(\mathrm{sSFR}/\mathrm{yr}^{-1}) \geq -11$ and quenched for $\log(\mathrm{sSFR}/\mathrm{yr}^{-1}) < -11$. Galaxies with SFR$=0$ are included in the quenched population. The resulting trends are consistent with the main-sequence-offset classification adopted in the main text, confirming the robustness of our results.
}
\label{fig:sSFR}
\end{figure*}

To test the sensitivity of our results to the classification adopted in Section~\ref{ms:ms_definition} (i.e Po19-based $\Delta\log(\mathrm{SFR})$), we repeated the analysis using a fixed threshold of $\log(\mathrm{sSFR}/\mathrm{yr}^{-1})=-11$ (see Fig.~\ref{fig:sSFR}). Although this method does not trace the entire SFR-$M_{\star}$ plane, it helps to assess the overall trend of star formation. 

Comparing the Po19-based $\Delta\log(\mathrm{SFR})$ (Fig.~\ref{fig:MS}, Table.~\ref{table:stats}) and sSFR (Fig.~\ref{fig:sSFR}) classifications yields a 10–20\% difference in quenched fractions at low stellar masses due to the green-valley population, but <5\% at higher masses. Crucially, the simulation ranking is unchanged: Magneticum shows the strongest quenching, followed by SIMBA and TNG100, confirming the robustness of our results.
\subsection{Centrals and satellites distribution}
\label{env:env_sect} 
Our sample includes centrals: [MaNGA (2,{}493), Box2 (453,{}081), Box4 (1,{}227), TNG100 (3,{}932, SIMBA (6,{892}), SDSS (298,{}440)] and satellites: [MaNGA (990), Box2 (201,{}707), Box4 (448), TNG100 (2,{}505), SIMBA (3,{}266), SDSS (62,{}888)]. Figure~\ref{fig:dist} shows that in both MaNGA, all simulations and the main SDSS sample, centrals increase with increase in $M_{\star}$, while it is vice versa for their counter parts satellites.

\section{Galaxy star-formation activity classification}
\label{bimodal:ms_bimodality}
As pointed by \cite{2019MNRAS.485.4817D}, the distinction between 'quenched' and 'star-forming' galaxies is inherently ambiguous and arbitrary. Both simulations and observations lack a non-arbitrary zero-point, as quenching typically represents either a numerical resolution limit or the sensitivity floor of star-formation tracers. 
The uncertainties and biases in the SFR measurements can create the appearance of a bimodal distribution \citep{2017MNRAS.470L..59F}. Therefore, in SFR-$M_{\star}$ plane, the terms bimodality, green valley and red sequence work as long as we count galaxies with unresolved/undetectable SFR (assigned zero SFR in simulations) and  log(sSFR)$<10^{-12}\,\mathrm{yr}^{-1}$ in observations, otherwise \cite{2017MNRAS.470L..59F, 2018MNRAS.473.3507E} the logarithmic distribution of SFR is unimodal forming a peak at the star-forming sequence and an extended tail towards low SFRs. 
\end{appendix} 

\end{document}